\documentclass[10pt,journal]{IEEEtran}
\usepackage{amsmath,amsfonts}
\usepackage{algorithmic}
\usepackage{array}
\usepackage[caption=false,font=normalsize,labelfont=sf,textfont=sf]{subfig}
\usepackage{textcomp}
\usepackage{stfloats}
\usepackage{url}
\usepackage{bm}
\usepackage{verbatim}
\usepackage{graphicx}
\usepackage{amsmath}
\usepackage{lineno,hyperref}
\usepackage{multirow}
\usepackage{booktabs} 
\usepackage{tabularx} 
\usepackage{caption}  
\usepackage{stfloats}
\def\BibTeX{{\rm B\kern-.05em{\sc i\kern-.025em b}\kern-.08em
    T\kern-.1667em\lower.7ex\hbox{E}\kern-.125emX}}
\usepackage{balance}
\begin{document}
\title{Dual-TSST: A Dual-Branch Temporal-Spectral- Spatial Transformer Model for EEG Decoding}
\author{Hongqi Li*~\IEEEmembership{Member,~IEEE,}, Haodong Zhang, Yitong Chen
\thanks{Manuscript received Aug 21, 2024. This work was supported in part by the Natural Science Basic Research Program of Shaanxi Province under Grant 2024JC-YBQN-0659, in part by the Guangdong Basic and Applied Basic Research Foundation under Grant 2022A1515110252, in part by the Basic Research Programs of Taicang under Grant TC2023JC16, in part by the Fundamental Research Funds for the Central Universities under Grant D5000210969 (Corresponding author: Hongqi Li.)

H. Li is with the School of Software, Northwestern Polytechnical University, Xi’an 710072, China, and also with the Research \& Development Institute of Northwestern Polytechnical University in Shenzhen, Shenzhen 518063, China, and also with the Yangtze River Delta Research Institute of Northwestern Polytechnical University, Taicang 215400, China (e-mail: lihongqi@nwpu. edu.cn)

H. Zhang and Y. Chen are with the School of Software, Northwestern Polytechnical University, Xi’an 710072, China (e-mail: zhang\_haodong@mail. nwpu.edu.cn, chenyt@mail.nwpu.edu.cn).}}

\markboth{Journal of \LaTeX\ Class Files,~Vol.~18, No.~9, September~2020}%
{How to Use the IEEEtran \LaTeX \ Templates}

\maketitle
\begin{abstract}
The decoding of electroencephalography (EEG) signals allows access to user intentions conveniently, which plays an important role in the fields of human-machine interaction. To effectively extract sufficient characteristics of the multichannel EEG, a novel decoding architecture network with a dual-branch temporal-spectral-spatial transformer (Dual-TSST) is proposed in this study. Specifically, by utilizing convolutional neural networks (CNNs) on different branches, the proposed processing network first extracts the temporal-spatial features of the original EEG and the temporal-spectral-spatial features of time-frequency domain data converted by wavelet transformation, respectively. These perceived features are then integrated by a feature fusion block, serving as the input of the transformer to capture the global long-range dependencies entailed in the non-stationary EEG, and being classified via the global average pooling and multi-layer perceptron blocks. To evaluate the efficacy of the proposed approach, the competitive experiments are conducted on three publicly available datasets of BCI IV 2a, BCI IV 2b, and SEED, with the head-to-head comparison of more than ten other state-of-the-art methods. As a result, our proposed Dual-TSST performs superiorly in various tasks, which achieves the promising EEG classification performance of average accuracy of 80.67\% in BCI IV 2a, 88.64\% in BCI IV 2b, and 96.65\% in SEED, respectively. Extensive ablation experiments conducted between the Dual-TSST and comparative baseline model also reveal the enhanced decoding performance with each module of our proposed method. This study provides a new approach to high-performance EEG decoding, and has great potential for future CNN-Transformer based applications.
\end{abstract}


\captionsetup{
  labelsep=space,     
}
\section{Introduction}
\label{chap:1}
\IEEEPARstart{B}{rain-Computer/Machine} interfaces (BCIS/BMIS) have garnered much attention over the past decades due to their outstanding ability to convert the users’ brain activity into machine-readable intentions or commands \cite{1}. Among various BCI modalities, noninvasive electroencephalograph (EEG) has the advantages of adequate temporal resolution, non-surgical electrode placements, and low cost, thus leading to its widest application in the fields of rehabilitation engineering \cite{2,3}, cognitive science \cite{4}, neuroscience, and psychology \cite{5}. 

Various brain paradigms, such as the motor imagery (MI), event-related P300, and steady-state visual evoked potentials (SSVEP), have been extensively studied by researchers \cite{5}, and a complete EEG-based BCI system generally consists of the user intention decoding process of the signal acquisition, preprocessing, feature extraction, classification, and a final application interface for the control signal convert. To gain an accurate interpreting of the sampled EEG, two main categories of recognition methods, traditional machine learning (ML) algorithms \cite{6} and advanced deep learning (DL) techniques \cite{7,8}, have been innovatively investigated. Traditional ML methods usually involve feature extraction and feature classifi-cation, where the former procedure uses algorithms of the common spatial pattern (CSP), Filter bank CSP, fast Fourier transform (FFT), wavelet transform, etc. As for the feature classification, the supervised learning approach (e.g., linear discriminant analysis (LDA), support vector machine (SVM)) and the unsupervised methods (e.g., K nearest neighbor (KNN)) have been shown to be effective. However, since features are manually extracted from the raw non-stationary EEG with low signal-to-noise ratio, the specific expertise is generally required, leading to the process being time-consuming and complicated. Worse more, the useful information may also be lost due to insufficient expert experience. The DL methods, on the other hand, allow end-to-end models that are composed of multiple processing layers to learn the data representation automatically, thereby minimizing the need for human manual intervention and domain-specific preprocessing, and have already achieved excellent even the state-of-the-art (SOTA) performance in several domains such as computer vision \cite{9,10}, and natural language processing \cite{11}.

Specifically for the EEG decoding, the convolutional neural networks (CNNs)-based ConvNet has reached comparable classification results to the traditional ML methods \cite{12}. A compact network called EEGNet has been proposed in \cite{13}, which utilized depth wise and separable convolutions to build an EEG-specific model that capable of learning features across various tasks. Moreover, a long short-term memory (LSTM) based recurrent neural network (RNN) has been developed by Tortora et al. in \cite{14} for decoding the gait events from EEG, where the network’s ability to handle the time-dependent information was fully leveraged. However, despite these commendable advances, CNNs and RNNs are not perfect in processing EEG signals. More specifically, while CNNs are good at learning local features, it is difficult to obtain long-term dependencies across the whole data scale. RNNs, on the other hand, are also prone to difficulties in capturing long-term dependencies in long sequence data. Therefore, to address these shortcomings, the research processing for sequence signals is gradually shifting to the self-attention mechanism, which allows each element in a sequence to be processed taking into account the relationship with all other elements, thus capturing richer contextual features. Moreover, the multi-feature analysis of EEG has also increasingly attracted attention considering that the sampled signals contain multi-dimensional features of the temporal, spectral, and spatial domains.
\vspace{-0.4cm}
\subsection{Related Work}
One of the most famous models based on the self-attention mechanism is the Transformer model, which has recently been attempted to EEG decoding. Sun et al. \cite{15} introduced a novel approach by integrating the multi-head attention mechanisms with CNNs for motor imagery tasks, and the various positional embedding techniques were used to improve the classification accuracy. As a result, the introduced five Transformer-based models have significantly outperformed existing models. Similarly, a compact hybrid model of CNNs and Transformers, named EEG Conformer, was developed to decode EEG signals by capturing both the local and global features, which was excelled on three public datasets and potentially established a new baseline for EEG processing \cite{16}. The proposed ADFCNN in \cite{17} utilized the convolutions at two different scales to capture comprehensive spatial details in EEG data, and the features were fused through a self-attention mechanism. Moreover, Arjun \cite{18}, Al-Quraishi \cite{19}, and Mulkey \cite{20} first converted the EEG data into time-frequency images, and then used Vision Transformers based on the idea of computer vision field. In the realm of pretrained models, different models of the BERT \cite{21}, GPT and Swin Transformer model \cite{22} have been designed to transform the EEG into textual and visual formats for the further processing. Particularly, a Transformer-like recognition approach of Speech2EEG has been proposed to leverage the pretrained speech processing networks for the robust EEG feature aggregation, thereby boosting EEG signal analysis capabilities \cite{23}. Given these above advancements, the promising potential of Transformers in EEG decoding has been well demonstrated.

On the other hand, with the development of DL models, a single feature can no longer satisfy the requirement of the performance improvement for increasingly complex models in EEG decoding, and therefore, multi-feature analysis methods gradually occupy the mainstream of EEG analysis methods. In 2019, Tian et al. \cite{24} crafted a multi-view DL strategy that first transforming the raw data into representations in the frequency and time-frequency domains, then independently extracting the features, which were finally merged to perform classification tasks efficiently. The data from multiple frequency bands was used in \cite{25} to create multi-view representations, where the spatial discrimination patterns of the views were learned by CNN, temporal information was aggregated by a variance layer, and the resultant features were classified by a fully connected layer. Recently, a multi-domain CNN model of TSFCNet was developed for MI decoding, which significantly outperformed the other traditional methods by extracting multi-scale features from the time domain and capturing the additional spatial, frequency, and time-frequency features \cite{26}. Earlier in this year, Liang et al. \cite{27} developed an EISATC-Fusion model to leverage the multi-scale EEG frequency band information combined with the attention mechanism and temporal convo¬lutional networks (TCN) for an integrated feature extraction process. In addition, a lightweight multi-feature attention CNN was proposed in \cite{28} to extract the information from frequency, localized spatial domains, and feature maps to enhance the precision of EEG analysis, where a hybrid neural network of SHNN was designed to autonomously extract the spatial, spectral, and temporal features from EEG \cite{29}. In conclusion of these mentioned studies, the research community has tended to extract the temporal-spatial-spectral features simultaneously, which helps to improve the understanding and decoding effects of specific EEG signals. However, since the EEG signals are first collected and expressed in temporal domain, while the frequency/time-frequency/spatial features are represented or converted by various approaches, how to efficiently extract and integrate the features from different dimensions and establish a more robust extraction process still remain challenging.
\vspace{-0.4cm}
\subsection{Contribution and Overview}
As mentioned earlier, the application of Transformer-based and multi-feature analysis in EEG decoding has just emerged and is in a phase of continuous development, and there are few attempts to naturally combine the two. Since the convolutional networks-based models are able to automatically learn more discriminative local features from raw EEG data, while the attention-based Transformer adepts to describe the long-range dependencies, the combination of these two modules is envisioned to benefit each other for a more comprehensive interpretation of human user EEG data. 

Driven by this insight, in present work, a novel decoding architecture model with dual-branch temporal-spectral-spatial transformer, termed as Dual-TSST, is proposed to extract multi-dimensional features hidden in EEG while considering their global correlations. Specifically, the proposed architecture mainly consists of three parts of the feature extraction, feature fusion, and classification modules. The first feature extraction module is composed of two branches of convolutional neural networks to receive multi-view inputs from raw EEG and to extract the inherent temporal-spectral-spatial features. These obtained features are fed into the feature fusion module to be jointly concatenated and then to learn their global relationships by a Transformer, and a classifier composed of multilayer perceptron and global pooling layers is finally used to achieve the results output. The main contributions of this study are summarized below.

1) We propose a natural fusion and collaboration architecture based on the classical CNNs and emerging Transformer, which is highly generalizable to a wide range of EEG decoding tasks. Specifically, the developed network enables to extract abundant powerful features without handcraft while allowing long-range correlation among features being considered and processed concurrently.

2) The designed Dual-TSST mainly comprises dual-scale convolution networks, wherein one is used to better extract the temporal feature from the raw EEG, and the other enables to acquire the time-frequency/time-spatial information from the converted EEG signals. These features are in the same scale to be concatenated and effectively jointly fused by a fusion patch. A self-attention mechanism is applied to adaptively enhance the flexibility of the feature fusion.

3) Dual-TSST has undergone extensive experiments on multiple public datasets to demonstrate the model structure and superior performance, the compared results with state-of-the- art models proved the effectiveness of the proposed method. 

The rest of this paper is organized as follows. Section \ref{chapt:2} introduces the design ideas and specific structural principles of Dual-TSST. Section \ref{chapt:3} presents the used datasets with related data preprocessing, and experimental setups. The comparable results and visualized model effects are presented in Section \ref{chapt:4}, while a detailed discussion and conclusion is given in the final Section \ref{chapt:5}.

\begin{figure*}[htpb]
\centering
\includegraphics[width=18cm]{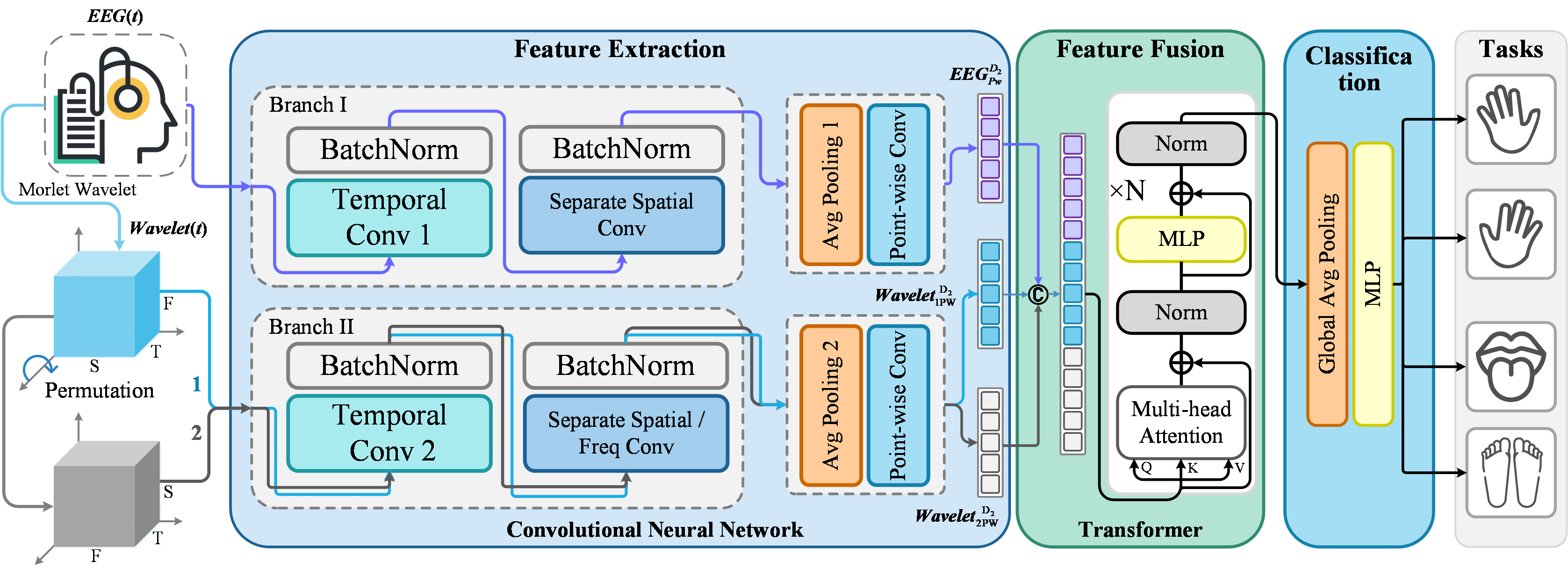}
\captionsetup{font=footnotesize}
\caption{The proposed Dual-TSST framework, including feature extraction of two CNN branches, feature fusion with Transformer, and classification modules.}
\vspace{-1.2\baselineskip} 
\label{fig_1}
\end{figure*}

\section{Approach of Dual-TSST Network}
\label{chapt:2}
The EEG signals are notable for their exceptional temporal resolution while encompassing extensive spectral and spatial properties. With the goal of processing EEG with multiple features involved being considered adequately and efficiently, a generalized network adheres to machine learning principles has been proposed, in which the advanced deep learning techniques are utilized to perform feature extraction, feature fusion, and the final classification step by step.
\vspace{-0.4cm}
\subsection{Overall Model Architecture}
Some traditional practice of EEG decoding generally uses exclusive raw EEG or solely time-frequency images derived from the transformed data, which may lead to leakage of contained information during the conversion process. Instead, as illustrated in Fig. \ref{fig_1}, a dual-branch model named Dual-TSST, capable of processing diverse views of EEG, is designed to start with both the given raw and converted EEG signals. For the first module of the feature extraction, two branches based on convolutional neural networks are applied to sufficiently extract potential characteristics from the temporal, frequency, and spatial domain. Branch II, in particular, is uniquely designed to simultaneously analyze the wavelet-transformed time-frequency EEG data in two separate flip-flop formats, collecting the time-frequency-space features comprehensively. To reduce the model complexity and computational load, the depth wise separable convolutions and average pooling layers are employed in this module.

The acquired features from branch I and branch II are then synergistically integrated and serve as the input of patch embedding for the feature fusion part, where a Transformer module is exploited to learn global relationships among the extracted properties. Ultimately, for the classification module, a global average pooling (GAP) layer and multilayer perceptron (MLP) module is used to analyze the inputted features and deliver the final classification outcomes. 
\vspace{-0.4cm}
\subsection{Source of Data}

1) Data Input: The original EEG data can be represented as $\textbf{\textit{EEG}}(t)\in \mathbb{R}^{ch\times T}$, where \textit{ch} is the number of electrodes indicating spatial dimensions, and \textit{T} represents the time samples of the EEG data. Initially, to convert the two-dimensional time-domain data into three-dimensional time-frequency domain, the Morlet wavelet transform \cite{30} provided by MNE-Python is applied, for which the process is described by

\vspace{-0.4cm}
\begin{equation}
   \textbf{\textit{W}}(a,t) = \frac{1}{\sqrt{a}} \int_{-\infty}^{+\infty} \textbf{\textit{EEG}}(\tau)\boldsymbol{\Psi}\left(\frac{\tau - t}{a} f_{o}\right) d\tau
\end{equation}

where \textbf{\textit{W}}(a,t) represents the transformed outcomes, a is the scale parameter related to frequency and sampling rate, fo means the central frequency, $\tau$ is the time variable, $\boldsymbol{\Psi}(t)$ is the wavelet function as 
\vspace{-0.1cm}
\begin{equation}
    \boldsymbol{\Psi}(t)=\frac{1}{\sqrt {\sigma \sqrt {\pi}} }e^{-t^{2}/\sigma _{t}^{2}}e^{i2\pi f_{o} t}
\end{equation}

and $\sigma_{t}$ is the wavelet’s temporal standard deviation.

For the Morlet wavelet transformation, we set the frequency \textit{freq} to match the frequency range used in the original data filtering. Noting that the number of cycles (i.e., n\_cycle) determines the width of the wavelet in the transformation, and is related to the temporal standard deviation $\sigma_{t}$. Larger n\_cycle results in a wider wavelet, leading to lower time but higher frequency resolution. Conversely, a smaller n\_cycle results in a narrower wavelet, enhancing both the time resolution and frequency resolution. To achieve a balance, n\_cycle is set to be half of the frequency, i.e., n\_cycle = \textit{freq}/2, and the sampling rate for the time resolution remains the same as that of the original EEG signal. We use  $\textbf{\textit{Wavelet}}(t) \in \mathbb{R}^{ch\times T\times F}$ to represent the transformed time-frequency data.

Then, both original $\textbf{\textit{EEG}}(t)$ and transformed time-frequency Wavelet data $\textbf{\textit{Wavelet}}(t)$ are subjected to Z-Score normalization, which preserves the data’s dimensional shape while ensuring consistency in analysis and can be represented as:
\vspace{-0.03cm}
\begin{equation}
    \textbf{\textit{x}}^{'}=\frac{\textbf{\textit{x}}-\mu }{\sigma }
\end{equation}
where $\boldsymbol{x}$ and $\boldsymbol{x}^{'}$ represents the input and output data, $\mu$ and $\sigma$ are the calculated mean and standard deviations.

2) Data Augmentation: To mitigate the challenge of limited EEG data availability in decoding, several data augmentation strategies can be applied. Here the Segment and Reassemble (S\&R) mechanism is adopted. More specifically, each EEG sample from the same category and its corresponding time-frequency data are divided into a predetermined number of fixed segments (labeled \textit{R}). These segments are subsequently reconnected in various random orders that respect the original temporal sequence. This technique not only diversifies the training dataset but also enhances the model’s ability to generalize from limited data samples. Following the guidelines set forth in references \cite{16,31}, we generated augmented data in each epoch, matching the batch size, thereby ensuring consistent model training across different data permutations.
\vspace{-0.4cm}
\subsection{Feature Extraction based on Dual CNN Branches}

\begin{figure}[htpb]
\centering
\includegraphics[width=8.8cm]{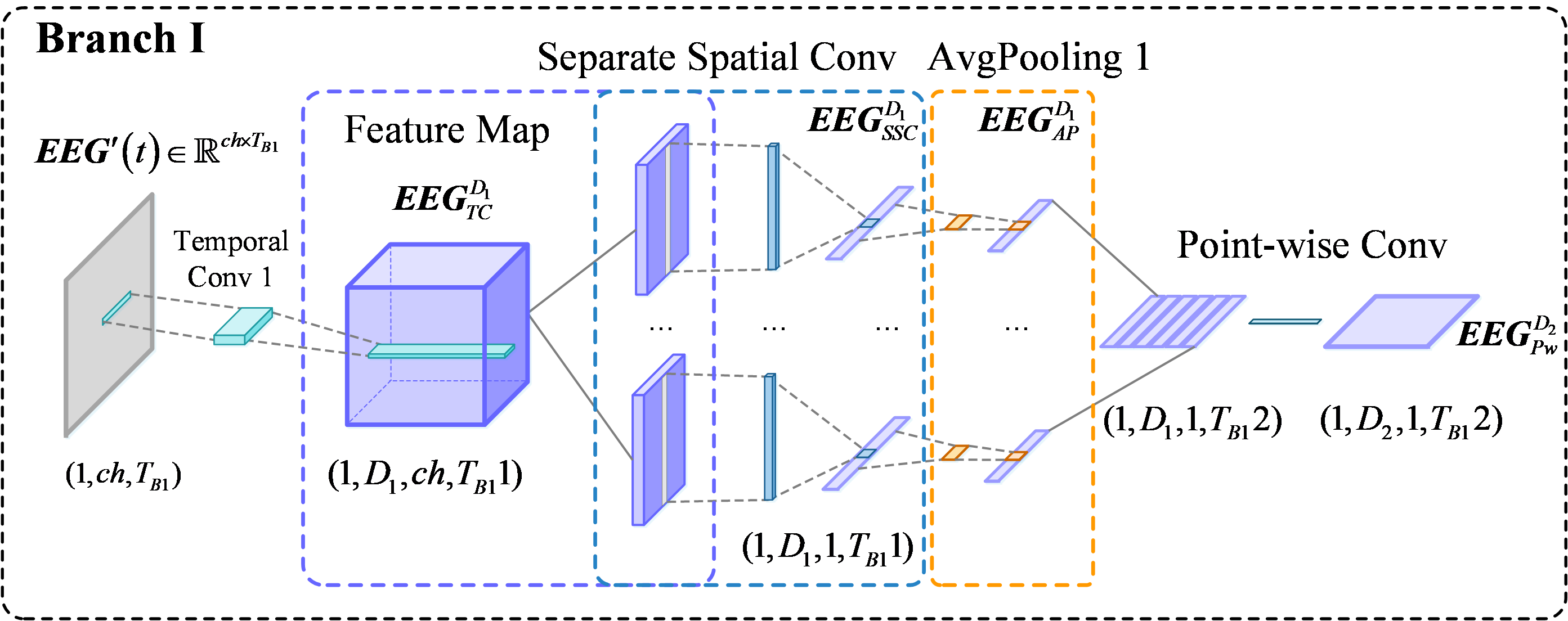}
\captionsetup{font=footnotesize}
\caption{Structure and data flow of Branch I for feature extraction module.}
\label{fig_2}
\end{figure}

1) Branch I for original $\textbf{\textit{EEG}}(t)$: As shown in Fig. \ref{fig_2}, the shape of the inputted 2D EEG data is $[ch \times T_{B1}]$. To extract features in the temporal dimension, the time convolution is first used, resulting in a 3D feature map of $\textbf{\textit{EEG}}_{TC}^{D_{1}}$, with the shape $[D_1 \times ch \times T_{B1}1]$. Here, to capture local details in the temporal dimension as much as possible, the time convolution kernel size is set to be relatively small. The relevant process can be summarized as:
\begin{equation}
    \textbf{\textit{EEG}}(t)_{TC}^{D_{1}}=TimeConv(\textbf{\textit{EEG}}^{'}) 
\end{equation}
Then, separable spatial convolution compresses the spatial dimension and extracts features from the electrodes, changing the feature map shape to $[D_1 \times 1 \times T_{B1}1]$. It should be noticed that, to ensure the performance, Batch Norm layers (see in Fig. \ref{fig_1}) are added after the time convolution and separable spatial convolution, and the ELU activation function is also employed. The above data flow can be expressed as follows:
\begin{equation}
    \textbf{\textit{EEG}}_{SSC}^{D_{1}}=ELU(BN(SSConv(\textbf{\textit{EEG}}_{TC}^{D_{1}})))
\end{equation}
where BN means the batch normalization function, and $SSConv$ indicates the related separable spatial convolution.

After that, an average pooling layer is used to extract features while reducing the data in the temporal dimension. With the enhanced generalization and noise suppressing ability, a feature map of $\textbf{\textit{EEG}}_{AP}^{D_{1}}$is derived, as the shape of $[D_1 \times 1 \times T_{B1}2]$. Finally, pointwise convolutions are applied for channel fusion and increasing the channel dimension to some extent, as enhancing the data’s information content and expressive power. The final feature map $\textbf{\textit{EEG}}_{Pw}^{D_{2}}$ is in shape of $[D_2 \times 1 \times T_{B1}2]$. The entire data flow of these mentioned operations can be represented by the following process:
\begin{equation}
    \textbf{\textit{EEG}}_{Pw}^{D_{w}}=PWConv(AP( \textbf{\textit{EEG}}_{SSC}^{D_{1}}))
\end{equation}

\vspace{-0.4cm}

2) Branch II for converted $\textbf{\textit{Wavelet}}(t)$ : Branch II is designed for processing the time-frequency EEG data. As illustrated in Fig. \ref{fig_1}, to capture multidimensional features, distinct inputs in different viewpoints (i.e., Input 1 and Input 2) are fed into such branch. Here, the Input 2 is obtained by subtly transposing Input 1 by 90 degrees. Unlike branch I of processing original EEG in a single line, branch II is actually perform simultaneous processing of multi-inputs. Specifically, the time convolution is applied first with a followed Batch normalization, which can be expressed as:
\begin{equation}
    {\textbf{\textit{Wavelet}}_i}_{TC}^{D_{1}}=TimeConv({\textbf{\textit{Wavelet}}_i}^{'}),i=1,2 
\end{equation}

where ${\textbf{\textit{Wavelet}}_i} (i=1,2)$ represents the branch inputs, and ${\textbf{\textit{Wavelet}}_i}_{TC}^{D_{1}}$ is the relevant output. 

\begin{figure}[htpb]
\centering
\includegraphics[width=8.8cm]{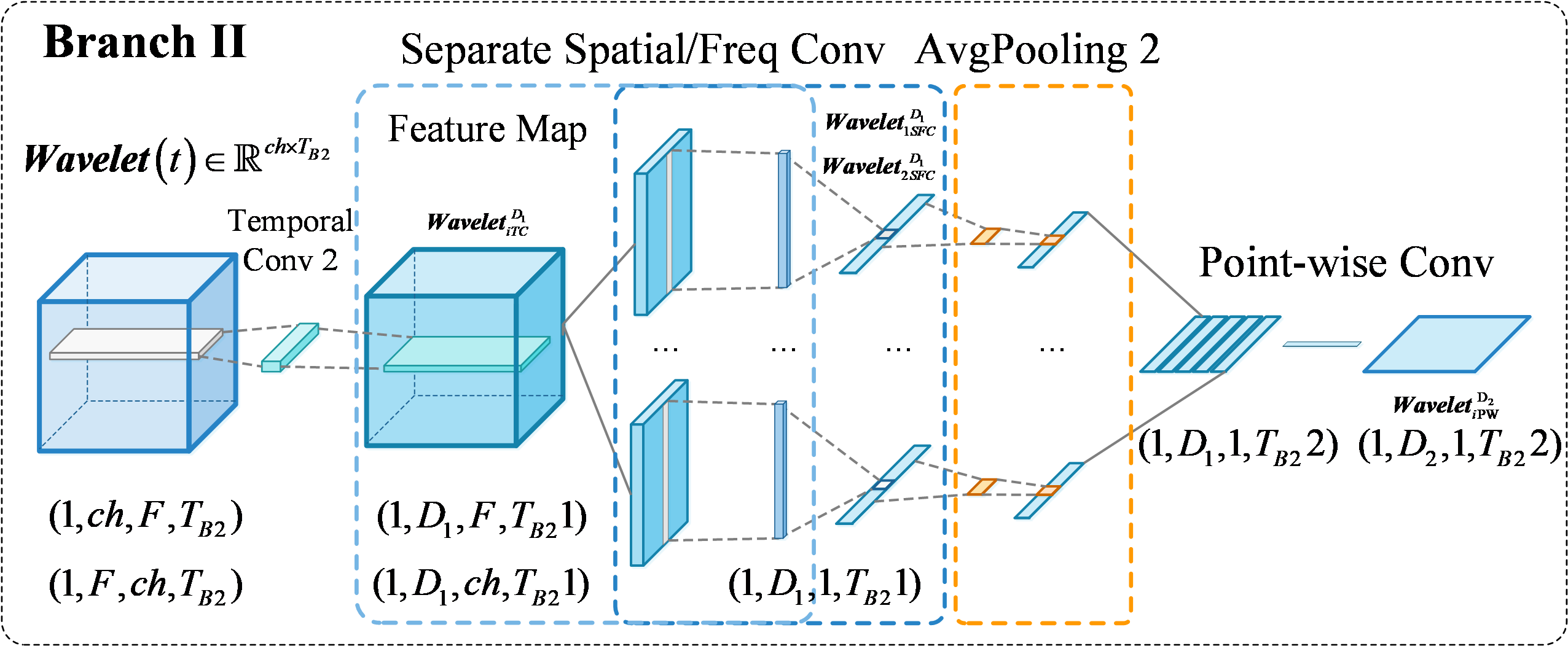}
\captionsetup{font=footnotesize}
\caption{Structure and data flow of Branch II for feature extraction module.}
\label{fig_3}
\end{figure}

Given the differences in temporal resolution and information content between the time-frequency $\textbf{\textit{Wavelet}}(t)$ and the original $\textbf{\textit{EEG}}(t)$, a different time convolution of scale is used, which has shapes $[ch\times F \times T_{B2}]$ and $[F\times ch \times T_{B2}]$, respectively. Indeed, as shown in Fig. \ref{fig_3}, such the choice is aimed to balance the features derived from the original and time-frequency data that will produce feature maps $[D_1\times F \times T_{B2}1]$ and $[D_1\times ch \times T_{B2}1]$.

Similarly, separable spatial and frequency convolutions are employed for feature extraction and dimension compression in the spatial and frequency dimensions, resulting in feature maps ${\textbf{\textit{Wavelet}}_1}_{SSC}^{D_{1}}$ and ${\textbf{\textit{Wavelet}}_2}_{SFC}^{D_{1}}$, both with shapes $[D_1\times 1 \times T_{B2}1]$. The detailed operation of these two processing are:

\vspace{-0.4cm}

\begin{equation}{\textbf{\textit{Wavelet}}_{1}}_{SSC}^{D_{1}}=ELU(BN(SSConv({\textbf{\textit{Wavelet}}_{1}}_{TC}^{D_{1}})))
    \tag{8a}
\end{equation}

\vspace{-0.6cm}
\begin{equation}{\textbf{\textit{Wavelet}}_{2}}_{SFC}^{D_{1}}=ELU(BN(SFConv({\textbf{\textit{Wavelet}}_{2}}_{TC}^{D_{1}})))
    \tag{8b}
\end{equation}

Subsequently, an average pooling layer is used to suppress noise, extract features, and reduce data volume, resulting in data with shape $[D_1\times 1 \times T_{B2}1]$. Finally, pointwise convolutions are applied to achieve channel fusion and dimension elevation, producing the feature maps ${\textbf{\textit{Wavelet}}_1}_{PW}^{D_{2}}$ and ${\textbf{\textit{Wavelet}}_2}_{PW}^{D_{2}}$, each with the shape $[D_2\times 1 \times T_{B2}1]$. Similarly, the hyper parameters D1 and D2 are set to be the same of branch I. The entire data flow of these descriptions is as follows:

\addtocounter{equation}{+1}
\vspace{-0.45cm}
\begin{equation}
    {\textbf{\textit{Wavelet}}_{i}}_{PW}^{D_{2}}=PWConv(AP( {\textbf{\textit{Wavelet}}_{i}}_{SSC/SFC}^{D_{1}}))
\end{equation}
\vspace{-0.95cm}
\subsection{Feature Fusion based on Transformer}
Three representative feature characteristics can be acquired from the above feature extraction process with Branch I and Branch II. To better integrate them, we reshape these outputs to be $\textbf{\textit{EEG}}_{S}^{D_{2}}$, ${\textbf{\textit{Wavelet}}_1}_{S}^{D_{2}}$, and ${\textbf{\textit{Wavelet}}_2}_{S}^{D_{2}}$ with shapes of $[T_{B1}2 \times D_2]$, $[T_{B2}2 \times D_2]$, and $[T_{B2}2 \times D_2]$, respectively. Such dimensional conversion is employed to suit the data need of the succeeding Transformer, which is applied to learn the cross-channel context information and the appropriate Encoder accepts inputs shaped as [SeqLength × FeatureSize]. The reshaped feature maps are horizontally concatenated to form a unified dataset $\textbf{EW}_{Fusion}$, which represents a fusion of the original EEG and time-frequency Wavelet data:

\vspace{-0.3cm}
\begin{equation}
    \textbf{EW}_{Fusion} = \text{Concat}(\textbf{\textit{EEG}}_{S}^{D_{2}}, 
    {\textbf{\textit{Wavelet}}_{1}}_{S}^{D_{2}}, {\textbf{\textit{Wavelet}}_{2}}_{S}^{D_{2}})
\end{equation}

The new feature $\textbf{EW}_{Fusion}$, which takes on the shape $[D_2\times T_{B1}2 \times T_{B2}2*2]$, is then processed using a multi-head attention mechanism within a complete Transformer Encoder. This setup captures the detailed correlations within the input sequence, thereby obtaining comprehensive global characteristics across time, space, and frequency dimensions from the combined EEG and time-frequency data.

An encoding approach akin to those in Vision Transformers is adopted, which involves parameterizable position encodings initialized with random values as:
\vspace{-0.1cm}
\begin{equation}
    \textbf{P} = Parameter(\textbf{P}_{init})
\end{equation}
\vspace{-0.5cm}
\begin{equation}
    \textbf{X}_{P} = \textbf{EW}_{Fusion} + \textbf{P}
\end{equation}
\vspace{-0.1cm}
where $\textbf{P}$ is the position encoding matrix, $\textbf{P}_{init}$ is its initial value, determined by random numbers, and $\textbf{X}_{P}$ represents the encoded feature matrix with shape of $[D_2\times T_{B1}2 \times T_{B2}2*2]$, which is subsequently mapped to the Query (\textit{Q}), Key (\textit{K}), and Value (\textit{V}) spaces through linear transformations, with learnable weighting matrices ${\textbf{\textit{W}}_{Q}}$, ${\textbf{\textit{W}}_{K}}$, and ${\textbf{\textit{W}}_{V}}$ as:
\vspace{-0.2cm}
\begin{equation}
    \textbf{\textit{Q}} = \textbf{\textit{X}}_{P}\textbf{\textit{W}}_{Q}, 
    \textbf{\textit{K}} = \textbf{\textit{X}}_{P}\textbf{\textit{W}}_{K}, 
    \textbf{\textit{V}} = \textbf{\textit{X}}_{P}\textbf{\textit{W}}_{V}
\end{equation}
\vspace{-0.5cm}
\begin{equation}
    Attention(Q,K,V)=Softmax(QK^T/\sqrt{D_2})V
\end{equation}
where $D_2$ is the dimensionality of patches within the data. 

The Transformer Encoder applies multi-head attention to parallelize the computation on data, thereby enhancing the expressivity and efficiency of the model and improving its generalizability. Multi-head attention (MHA) includes several self-attention layers, where each head generates an attention output, and the outputs from all heads are concatenated to form the final multi-head attention, as depicted in the following:

\vspace{-0.5cm}
\begin{equation}
    MHA(Q,K,V) = Concat(head_1,...head_h)W^O
\end{equation}
\vspace{-1.2cm}

\begin{equation}
   Head_i = SelfAttention(QW^Q_i,KW^K_i,VW^W_i)
\end{equation}

where $h$ denotes the number of heads, $W^o$ is the weight matrix that integrates information captured by different heads of $head_{i}$.

After the multi-head attention mechanism, as can be seen in Fig. \ref{fig_1}, a series of residual connections and layer normalization are performed to facilitate the information flow and stabilize the training process. The output is further processed using the MLP, followed by additional layer normalization, and residual connections, culminating in the final outputs from the multiple Encoder layers.
\vspace{-0.4cm}
\subsection{Classification Module}

The outcome of the Transformer Encoder maintains the same dimensional structure as its input. To effectively distill this complex data, global average pooling (GAP) is employed, which simplifies the feature map by averaging out the features over the entire spatial extent of each channel. This process extracts pivotal global information that is crucial for the next stage of processing.

Following the pooling, the data is routed to an MLP module with two linear layers. The Softmax function, which normalizes the linear outputs to form a probability distribution over the predicted output classes, aids in the transformation of the pooled features into an M-dimensional vector. The model’s performance is evaluated using a cross-entropy loss function, which is essential for classification tasks and is mathematically represented as:
\begin{equation}
    l=-\frac{1}{N_b}\sum_{i=1}^{N_b}\sum_{j=1}^{N_c}ylog(\hat{y})
\end{equation}
where $N_b$ is the batch size indicating the number of samples processed per training iteration, $N_c$  denotes the total number of categories in the classification task, $y$ is the true label of the data, and $\hat{y}$ is the predicted probability for each class. Briefly, the function   can effectively measures the difference between the predicted probabilities and the actual distribution, guiding the model towards more accurate predictions through training.
\vspace{-0.3cm}
\section{Dataset and Experimental Setup}
\label{chapt:3}
To evaluate the proposed method, we utilized three public datasets. Specifically, two BCI competition datasets in MI \cite{32}, sourced from MOABB (Mother of All BCI Benchmarks) project \cite{33}, and one widely used emotional SEED dataset \cite{34} are included. This section gives the relevant introduction and several required necessary procedures.
\vspace{-0.4cm}
\subsection{Datasets}

\textbf{\textit{Dataset I}}: \textbf{BCI Competition IV 2a} - This dataset comprises EEG recordings from 9 subjects that performing four distinct MI tasks, i.e., imagery movements of the left hand, right hand, both feet, and the tongue. Data were collected by 22 positioned Ag/AgCl electrodes according to international 10-20 system. To ensure signal quality, a 250 Hz sampling rate was utilized and the recorded data was filtered between 0.5 Hz and 100 Hz. The dataset includes two sessions, where the first session serves as the training set, and the second as the test set. Each session consists of six runs, with 48 trials per run that distributed evenly across task categories. In our study, we set the time window for each trial of this dataset between 2 and 6 seconds, and filtered the data using a frequency range from 0 to 40 Hz.

\textbf{\textit{Dataset II}}: \textbf{BCI Competition IV 2b} - This dataset features by the data from 9 subjects engaged in left and right hand MI tasks. The recording data were captured from three electrodes of C3, Cz, and C4, with a sampling frequency of 250 Hz. A band-pass filter of 0.5-100 Hz and a notch filter at 50 Hz have been used. Each subject participated in five sessions, where the initial two collected data without visual feedback and the subse¬quent three sessions included online feedback. Moreover, the dataset designates the initial three sessions (400 trials in total) for training and the final two (i.e., 320 trials) for testing. Noting that in our study each trial is allocated a time window from 3 to 7.5 s, with data similarly filtered within the 0 to 40 Hz range.

\textbf{\textit{Dataset III}}: \textbf{SEED} - Provided by BCMI Lab from Shanghai Jiao Tong University, this dataset consists of EEG data from 15 subjects who viewed clips from Chinese films edited to evoke various emotions (e.g., positive, negative, neutral). The films last for 4 minutes, with data processed using 1-seconds or 4s sliding windows across 62 channels, and downsampled to 200 Hz. Each subject underwent three experimental sessions, with data filtered through a 0-75 Hz band-pass filter. In addition, five-fold/ten-fold cross-validation techniques were involved in training. Also, a band-pass filter ranging from 0.5 Hz to 50 Hz was utilized on the SEED dataset, and the continuous data from each experiment was segmented into 1-second windows. 
\vspace{-0.9cm}
\subsection{Experiment Setting}
We constructed the developed model using Python 3.11 and PyTorch 2.0, and conducted training on a Nvidia GeForce RTX 4090 GPU using the Adam optimizer. The Adam optimizer was configured with a learning rate of 0.0001 and a weight decay of 0.0012, with $\beta_1$ and $\beta_2$ values at 0.5 and 0.999, respectively. Throughout the training, the epoch value was set to 1000, with a batch size of 32. The critical hyperparameters D1 and D2 were set to 40 and 120. On Datasets I and II, the data augmentation parameters R were designated as 8 and 9. Since the data scale is enough, no data augmentation was applied in Dataset III. The learning rate was adjusted with Cosine Annealing \cite{35}, which can be explained by the following formula:
\vspace{-0.1cm}
\begin{equation}
    lr = lr_{min} + 1/2(lr_{max} - lr_{min})(1+cos\frac{T_{cur}}{T_{max}}\pi)
\end{equation}

where $lr$ is the current learning rate, $lr_{max}$ and $lr_{min}$ are the related maximum and minimum values, respectively. $T_{cur}$ is the current training epoch, and $T_{max}$ is the total number of training epochs in a cycle. The learning rate decreases to $lr_{min}$ at the end of a cycle. For the experiments, $T_{max}$ was set to 32 to allow better model convergence and generalization during training.
\vspace{-0.4cm}
\subsection{Choice of Model Parameters}
Table \ref{tab:table 1} illustrates the input shapes, kernels, strides, and output configurations for each layer in the feature extraction, emphasizing how each layer contributes to the final outputs.
\begin{table}[htpb]
	\captionsetup{font=footnotesize,justification=centering}
    \caption{\textsc{\\Different Features Extracted on EEG by Basic Transformer Models}\label{tab:table 1}}
    \footnotesize
    \centering
	\begin{tabularx}{\columnwidth}{>{\centering\arraybackslash}p{0.1cm} >{\centering\arraybackslash}p{0.1cm} >{\centering\arraybackslash}p{0.1cm} >{\centering\arraybackslash}p{0.1cm}>{\centering\arraybackslash}p{0.7cm} >{\centering\arraybackslash}p{1.7cm} >{\centering\arraybackslash}p{0.6cm} >{\centering\arraybackslash}p{0.4cm} >{\centering\arraybackslash}p{1.3cm}}
		\toprule
        \multicolumn{4}{c}{\textbf{Module}}& \textbf{Layer*} & \textbf{Input shaped} & \textbf{Kernel} & \textbf{Stride} & \textbf{Output}\\
        \midrule
		\multicolumn{4}{c}{\multirow{4}*{Branch I}}  & TC & ($ch,T$) & (1,30) & (1,1) & ($D_1,ch,T_1$) \\
		\multicolumn{4}{c}{~}   & SSC & ($D_1,ch,T_1$) & ($ch,1$) & (1,1) & ($D_1,1,T_1$)\\
        \multicolumn{4}{c}{~}   & AP & ($D_1,1,T_1$)) & (1,120) & (1,12) & ($D_1,1,T_2$)\\
        \multicolumn{4}{c}{~}   & PWC & ($D_1,1,T_2$) & (1,1) & (1,1) & ($D_2,1,T_2$)\\
        \midrule
        \multicolumn{2}{c}{\multirow{8}*{Branch II}} & \multicolumn{2}{c}{\multirow{4}*{1}}  & TC & ($ch,F,T$) & (1,125) & (1,1) & ($D_1,F,T_1$) \\
		\multicolumn{2}{c}{~}   & \multicolumn{2}{c}{~}   & SFC & ($D_1,F,T_1$) & ($F,1$) & (1,1) & ($D_1,1,T_1$)\\
        \multicolumn{2}{c}{~}   & \multicolumn{2}{c}{~}   & AP & ($D_1,1,T_1$) & (1,64) & (1,32) & ($D_1,1,T_2$)\\
        \multicolumn{2}{c}{~}   & \multicolumn{2}{c}{~}   & PWC & ($D_1,1,T_2$) & (1,1) & (1,1) & ($D_2,1,T_2$)\\
        \cmidrule{3-9}
        \multicolumn{2}{c}{~} & \multicolumn{2}{c}{\multirow{4}*{2}}  & TC & ($F,ch,T$) & (1,125) & (1,1) & ($D_1,ch,T_1$) \\
		\multicolumn{2}{c}{~}   & \multicolumn{2}{c}{~}   & SSC & ($D_1,ch,T_1$) & ($ch,1$) & (1,1) & ($D_1,1,T_1$)\\
        \multicolumn{2}{c}{~}   & \multicolumn{2}{c}{~}   & AP & ($D_1,1,T_1$) & (1,64) & (1,32) & ($D_1,1,T_2$)\\
        \multicolumn{2}{c}{~}   & \multicolumn{2}{c}{~}   & PWC & ($D_1,1,T_2$) & (1,1) & (1,1) & ($D_2,1,T_2$)\\
    \bottomrule
	\end{tabularx}
	\caption*{\scriptsize * TC: Time Convolution, SSC: Separate Spatial Convolution, AP: Average Pooling, PWC: Point-wise Convolution, SFC: Separate Frequency Convolution}
 \vspace{-0.5cm}
\end{table}

Specifically, from the model structure, it is apparent that the final feature size outputted by each branch is primarily governed by the kernel size and stride of the Average Pooling layer. For Branch I, a relative small convolution kernel is set to capture more granular features along the temporal dimension. However, despite richer details can be extracted, it may result in a larger feature map size. Using a larger Pooling Kernel size helps control the map size and the receptive field of the features. Meanwhile, it helps to reduce the computational requirements and enhance the model’s generalization capabilities while maintaining substantial contextual information. In contrast, a larger convolution kernel set in Branch II aims to capture broader features along the time-frequency dimension, and a following smaller Pooling Kernel Size may facilitate more intensive feature extraction. Indeed, balancing the convolution kernel sizes and pooling parameters between different branches enhances the model’s flexibility, which helps to better adapt to the model’s intrinsic structure and allow the model to learn features of different scales from different data types, thus improving model performance. Here, to balance the features obtained while enhancing the model performance, we set a larger Pooling Kernel size $P_1$ of 120 with a stride of $P_1/10$ for Branch I, and a smaller Pooling Kernel size $P_2$ of 64 with a stride of $P_2/2$ for Branch II.

Moreover, the Transformer Encoder was configured with 4 blocks, and the multi-head attention mechanism was set with 10 heads. Finally, the model’s performance was evaluated using classification accuracy and the Kappa value, and the Kappa value is defined as:
\begin{equation}
    Kappa=\frac{P_o-P_e}{1-P_e}
\end{equation}
where $P_o$ is the proportion of correctly classified samples to the total number of samples, i.e., overall classification accuracy, and $P_e$ represents the probability of chance agreement, i.e., the correctness of random guesses.

Besides, we also used the Wilcoxon Signed-Rank Test to analyze the potential statistical significance.

\begin{table*}[!b]
    \captionsetup{font=footnotesize,justification=centering}
    \caption{\textsc{\\Comparison Results Of Different Methods On Dataset I [Avg Acc: The Average Accuracy(\%)}\label{tab:table 2}}
    \centering
    \footnotesize
    \resizebox{\textwidth}{!}{
    \begin{tabular}{cccccccccccccc}
        \toprule
        \textbf{Year} & \textbf{Methods} & \textbf{S1} & \textbf{S2} & \textbf{S3} & \textbf{S4} & \textbf{S5} & \textbf{S6} & \textbf{S7} & \textbf{S8} & \textbf{S9} & \textbf{Avg Acc} & \textbf{Std} & \textbf{Kappa} \\
        \midrule
        2017 & ConvNet \cite{12}    & 76.39 & 55.21 & 89.24 & 74.65 & 56.94 & 54.17 & 92.71 & 77.08 & 76.39 & 72.53 & 13.43 & 0.6337 \\
        2018 & EEGNet \cite{13}     & 85.76 & 61.46 & 88.54 & 67.01 & 55.90 & 52.08 & 89.58 & 83.33 & 86.87 & 74.50 & 14.36 & 0.66 \\
        2021 & FBCNet \cite{36}     & 85.42 & 60.42 & 90.63 & 76.39 & 74.31 & 53.82 & 84.38 & 79.51 & 80.90 & 76.20 & 11.28 & 0.6827 \\
        2021 & DRDA \cite{37}       & 83.19 & 55.14 & 87.43 & 75.28 & 62.29 & 57.15 & 86.18 & 83.61 & 82.00 & 74.74 & 12.22 & 0.6632 \\
        2022 & SHNN \cite{29}       & 82.76 & 68.97 & 79.31 & 65.52 & 58.62 & 48.28 & 86.21 & \textbf{89.66} & \textbf{89.87} & 74.26 & 13.93 & 0.6648 \\
        2022 & DAFS \cite{38}       & 81.94 & 64.58 & 88.89 & 73.61 & 70.49 & 56.60 & 85.42 & 79.51 & 81.60 & 75.85 & 9.86  & 0.678 \\
        2022 & EEG-ITNet \cite{39}  & 84.38 & 62.85 & 89.93 & 69.10 & 74.31 & 57.64 & 88.54 & 83.68 & 80.21 & 76.74 & 10.82 & - \\
        2023 & IFNet \cite{40}      & 88.47 & 56.35 & 91.77 & 73.78 & 69.72 & 60.42 & 89.24 & 85.42 & 88.72 & 78.21 & 12.73 & - \\
        2023 & Conformer \cite{16}  & 88.19 & 61.46 & 93.40 & \textbf{78.13} & 52.08 & 65.28 & 92.36 & 88.19 & 88.89 & 78.66 & 14.42 & 0.7155 \\
        2024 & ADFCNN \cite{17}     & 87.15 & 61.45 & \textbf{93.75} & 75.69 & 75.34 & 65.27 & 88.54 & 82.29 & 85.06 & 79.39 & 10.23 & - \\
        2024 & M-FANet \cite{28}    & 86.81 & \textbf{75.00} & 91.67 & 73.61 & 76.39 & 61.46 & 85.76 & 75.69 & 87.15 & 79.28 & \textbf{8.84}  & 0.7259 \\
        2024 & \textbf{Dual-TSST}       & \textbf{91.32} & 59.38 & 93.40 & 69.44 & \textbf{77.79} & \textbf{68.75} & \textbf{94.44} & 85.76 & 85.76 & \textbf{80.67} & 11.76 & 0.7413 \\
        \bottomrule
    \end{tabular}
    }
\end{table*}

\section{Results}
\label{chapt:4}
In this section, we compared the relevant results of proposed model against a variety of innovative state-of-the-art methods, where ablation experiments also being involved to demonstrate the contribution of each model part. Finally, we illustrated the interpretability of the extracted features visually, which helps in understanding underlying principles of the model.
In general, for the selected datasets, the classification effect of existing deep learning has made great progress compared to machine learning. To avoid redundancy and ensure the persua¬siveness of the comparison, we mainly review the latest methods of the relevant datasets in the past three years, as well as some well-established deep learning techniques (e.g., ConvNet \cite{12}, EEGNet \cite{13}, FBCNet \cite{36}, EEG Conformer \cite{16}). Among used methods, DRDA \cite{37} offers a sophisticated end-to-end domain adaptation approach tailored for EEG-based motor imagery classification tasks, and DAFS \cite{38} merges small sample learning with domain adaptation, enhancing domain-specific classification efficacy in MI-EEG tasks by leveraging source domain insights. EEG-ITNet \cite{39} features an interpretable CNN framework that relies on inception modules and dilated causal convolutions, whereas IFNet \cite{40} is a streamlined interactive convolutional network focusing on the interplay among various frequency signals to boost EEG feature depiction. MANN in \cite{41} integrates multiple attention mechanisms with transfer learning for EEG classification, and incorporates domain adaptation techniques to enhance its efficacy, while a multi-scale hybrid convolutional network of MSHCNN \cite{42} leverages convolu-tions across different dimen¬sions to distinctly extract temporal and spatial features from EEG data. In addition, several other latest models such as TSFCNet \cite{26}, Speech2EEG \cite{23}, EISATC-Fusion \cite{27}, FSA- TSP \cite{43}, FTCN \cite{46} have also been used for the evaluation.

\begin{table*}[htbp]
    \captionsetup{font=footnotesize,justification=centering}
    \caption{\textsc{\\Comparison Results Of Different Methods On Dataset II [Avg Acc: The Average Accuracy(\%)}
    \label{tab:table 3}}
    \centering
    \footnotesize
    \resizebox{\textwidth}{!}{
    \begin{tabular}{cccccccccccccc}
        \toprule
        \textbf{Year} & \textbf{Methods} & \textbf{S1} & \textbf{S2} & \textbf{S3} & \textbf{S4} & \textbf{S5} & \textbf{S6} & \textbf{S7} & \textbf{S8} & \textbf{S9} & \textbf{Avg Acc} & \textbf{Std} & \textbf{Kappa} \\
        \midrule
        2017 & ConvNet \cite{12}    & 76.56 & 50.00 & 51.56 & 96.88 & 93.13 & 85.31 & 83.75 & 91.56 & 85.62 & 79.37 & 16.27 & 0.5874 \\
        2018 & EEGNet \cite{13}     & 75.94 & 57.64 & 58.43 & 98.13 & 81.25 & 88.75 & 84.06 & 93.44 & 89.69 & 80.48 & 13.63 & 0.6096 \\
        2021 & DRDA \cite{37}       & 81.37 & 62.86 & 63.63 & 95.94 & 93.56 & 88.19 & 85.00 & 95.25 & 90.00 & 83.98 & 11.94 & 0.6796 \\
        2022 & MANN \cite{41}       & 82.81 & 60.36 & 59.06 & 97.50 & 91.88 & 86.38 & 84.06 & 93.44 & 86.88 & 82.54 & 12.95 & 0.6510 \\
        2022 & SHNN \cite{29}       & 83.33 & 61.76 & 58.33 & 97.30 & 91.89 & 88.89 & 86.11 & 92.11 & 91.67 & 83.49 & 13.10 & 0.6697 \\
        2023 & Conformer \cite{16}  & 82.50 & 65.71 & 63.75 & \textbf{98.44} & 86.56 & 90.31 & 87.81 & 94.38 & \textbf{92.19} & 84.63 & 11.49 & 0.6926 \\
        2023 & TSFCNet \cite{26}    & 76.25 & 70.00 & 83.75 & 97.50 & 72.81 & 86.56 & 88.44 & 92.50 & 89.69 & 86.39 & 8.81  & 0.7234 \\
        2023 & MSHCNN \cite{42}     & \textbf{86.80} & \textbf{77.94} & 65.97 & 97.97 & 93.24 & 88.88 & 86.80 & 82.89 & 86.80 & 85.25 & 8.67  & - \\
        2023 & Speech2EEG \cite{23} & 80.70 & 62.04 & 71.74 & 96.09 & 94.51 & 84.06 & 84.06 & \textbf{95.65} & 87.76 & 84.07 & 10.80 & - \\
        2024 & ADFCNN \cite{17}     & 79.37 & 72.50 & 82.81 & 96.25 & \textbf{99.37} & 84.68 & 93.43 & 95.31 & 86.56 & 87.81 & \textbf{8.39}  & - \\
        2024 & EISATC-Fusion \cite{27} & 75.00 & 72.86 & \textbf{86.56} & 96.88 & 97.81 & 84.38 & \textbf{94.06} & 93.75 & 86.88 & 87.58 & 8.54  & 0.7515 \\
        2024 & \textbf{Dual-TSST}            & 85.63 & 66.20 & 84.06 & 98.13 & 98.44 & \textbf{90.94} & 89.06 & 93.13 & 92.19 & 88.64 & 9.17  & \textbf{0.7718} \\
        \bottomrule
    \end{tabular}
    }
\end{table*}

\begin{table}[htbp]
    \small 
    \captionsetup{font=footnotesize,justification=centering}
    \caption{\textsc{\\Comparison Results Of Different Methods On Dataset III}\label{tab:table 4}}
    \centering
    \footnotesize
    \begin{tabularx}{\columnwidth}{>{\centering\arraybackslash}p{1cm} >{\centering\arraybackslash}X>{\centering\arraybackslash}p{1.2cm}>{\centering\arraybackslash}p{1cm}>{\centering\arraybackslash}p{1cm}} 
        \toprule
        \textbf{Year} & \textbf{Methods} & \textbf{Avg Acc} & \textbf{Std} & \textbf{Kappa} \\
        \midrule
        -    & SVM          & 80.80  & 12.87 & - \\
        2021 & BiHDM \cite{44} & 93.12  & 6.06  & - \\
        2022 & RGNN \cite{45}  & 79.37  & 10.54 & - \\
        2022 & EeT \cite{47}   & 96.28  & 4.39  & - \\
        2023 & FSA-TSP \cite{43} & 93.55  & 5.03  & - \\
        2023 & Conformer \cite{16} & 95.30  & -     & 0.9295 \\
        2024 & FTCN \cite{46}  & 89.13  & 4.49  & - \\
        2024 & \textbf{Dual-TSST}       &\textbf{96.65}  & \textbf{1.93}  & \textbf{0.9488} \\
        \bottomrule
    \end{tabularx}
    \vspace{-0.5cm}
\end{table}

\vspace{-0.4cm}

\subsection{Head-to-head Comparison Results}

Table \ref{tab:table 2} lists the comparison results of different algorithms applied in Dataset I. Specifically, the proposed Dual-TSST outperformed the existing SOTA methods in terms of overall average classification accuracy and the Kappa metrics, notably across subjects S1, S5, S6, and S7. In particular, compared with classical EEG decoding techniques like ConvNet and EEGNet, the average accuracy under current model has improved by 8.14\% (p\textless 0.05) and 6.17\% (p\textless 0.05), respectively, also with an obviously corresponding rise in Kappa values. Such the results underscore Dual-TSST’s enhanced capability for global feature extraction, as opposed to those local feature focus seen in ConvNet and EEGNet. Moreover, for the most test subjects, Dual-TSST achieved superior results over the FBCSP-inspired FBCNet and domain adaptation methods like DRDA and DAFS (p\textless 0.05), although being slightly inferior on S2 and S4. Compared to the models that introduced attention mechanisms into deep learning networks, such as Conformer, ADFCNN, and M-FANet, the developed Dual-TSST also showed better performance in most subjects’ accuracy, the average classifi¬cation accuracy, and the Kappa value. Among all the compared methods, the SHNN model excels in subjects S8 and S9, while being less accurate in S6. Overall, our proposed dual-TSST framework delivers varied improvements in the classification accuracy among different subjects within the Dataset I, and leads in terms of the average accuracy, and the Kappa metrics.

For the binary classification Dataset II, as we see from Table \ref{tab:table 3}, several additional models have been supplemented for the evaluation. Consequently, near the similar effects have been observed with those in Dataset I, where the dual-TSST not only surpasses conventional deep learning models such as ConvNet and EEGNet but significantly outperformed other advanced methods like DRDA, SHNN, Conformer, and ADFCNN, in almost all metrics (p\textless 0.05). In head-to-head comparisons with other leading techniques of MANN, TSFCNet, MSHCNN, and EISATC-Fusion, Dual-TSST consistently achieved superior average accuracy and Kappa values. Furthermore, the standard deviation values of the proposed dual-TSST in Table \ref{tab:table 3} is 9.17, which is relatively lower to most of the compared methods. Such the result underscores the model’s robust generalization ability to deliver steady strong results for diverse subjects.

\vspace{-0.5\baselineskip} 
\begin{figure}[h]
\centering
\captionsetup[subfloat]{font=scriptsize,labelformat = empty}
\subfloat[][]{\includegraphics[width = 7cm]{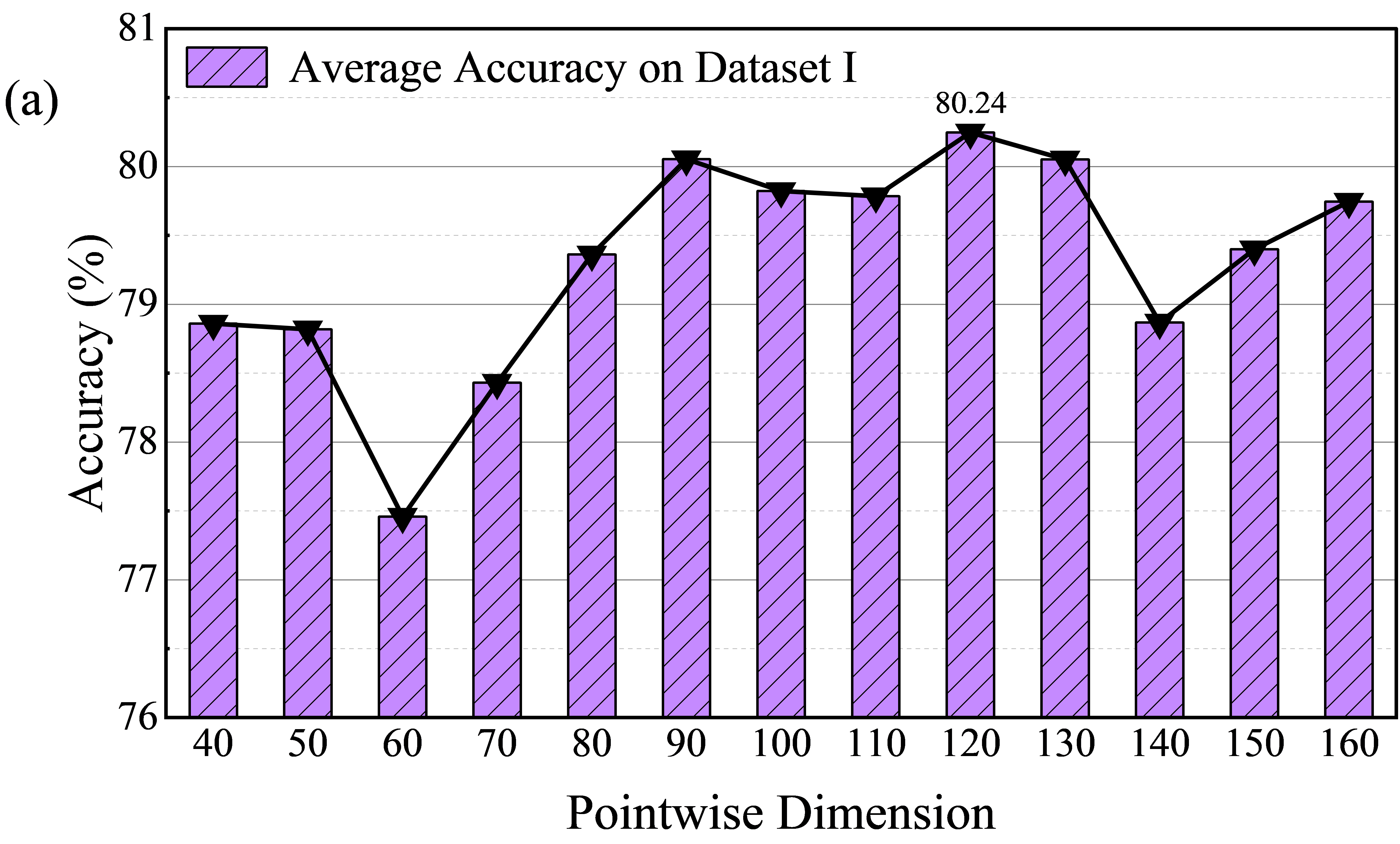}%
\label{fig_4 a}}\\
\vspace{-0.5cm}
\subfloat[][]{\includegraphics[width = 7cm]{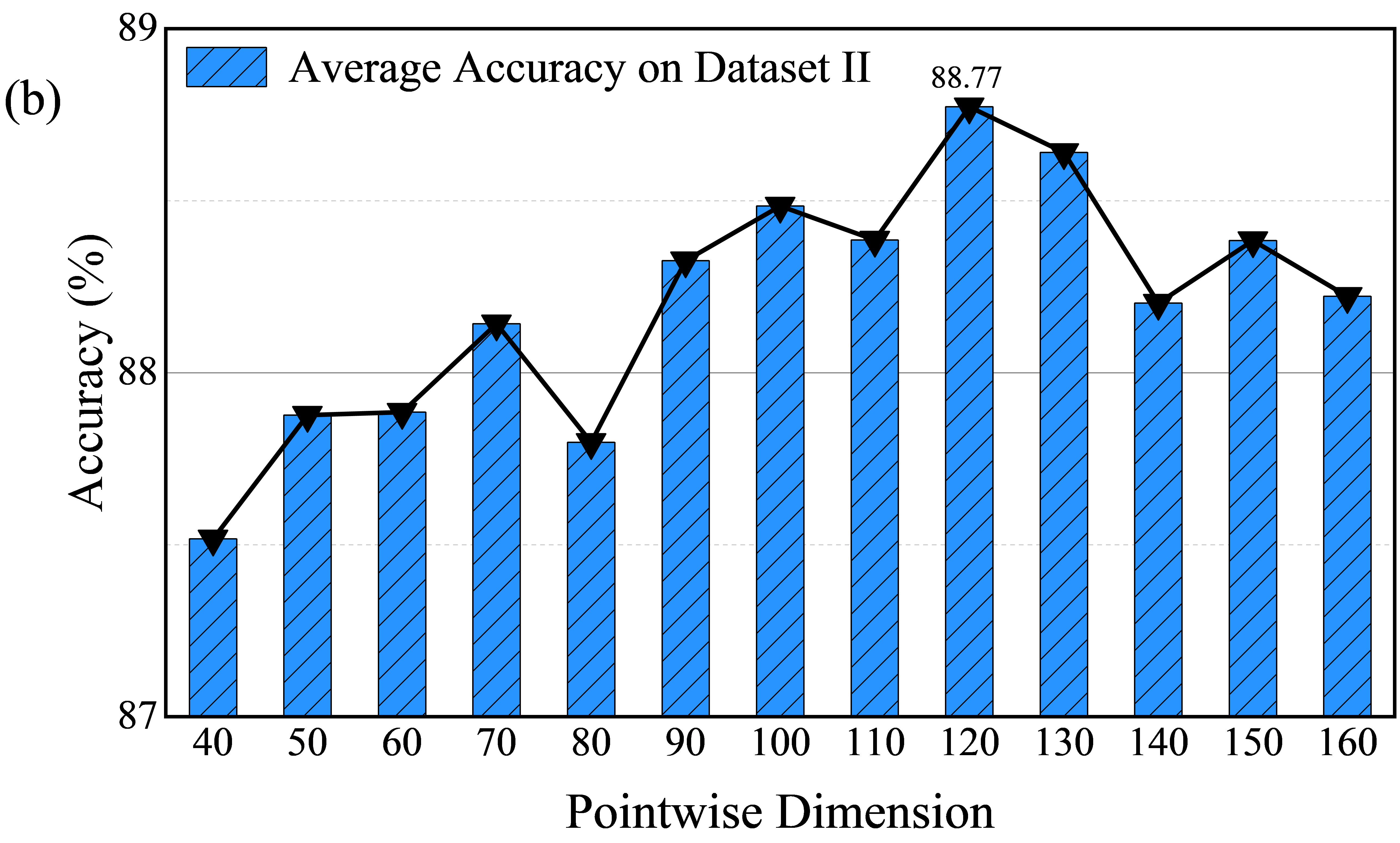}%
\label{fig_4 b}}
\captionsetup{font=footnotesize}
\vspace{-0.3cm}
\caption{The influence of pointwise dimension on model performance.}
\label{fig_4}
\end{figure}

To further evaluate the robustness and generalization ability of the model, we extended our analysis with the challenging emotion Dataset III of SEED, which presents a different type of task and requires the model to adapt to new patterns. As listed in Table \ref{tab:table 4}, the model continues to outperform the traditional machine learning algorithms and majority of the compared SOTA methods, indicating a commendable level of adapta¬bility of the designed model to effectively capture and interpret complex patterns associated with widely used EEG paradigms.

\vspace{-0.3cm}
\subsection{Parameter Sensitivity}

\vspace{-0.3cm}
\begin{figure}[hbtp]
\centering
\includegraphics[width=7.62cm]{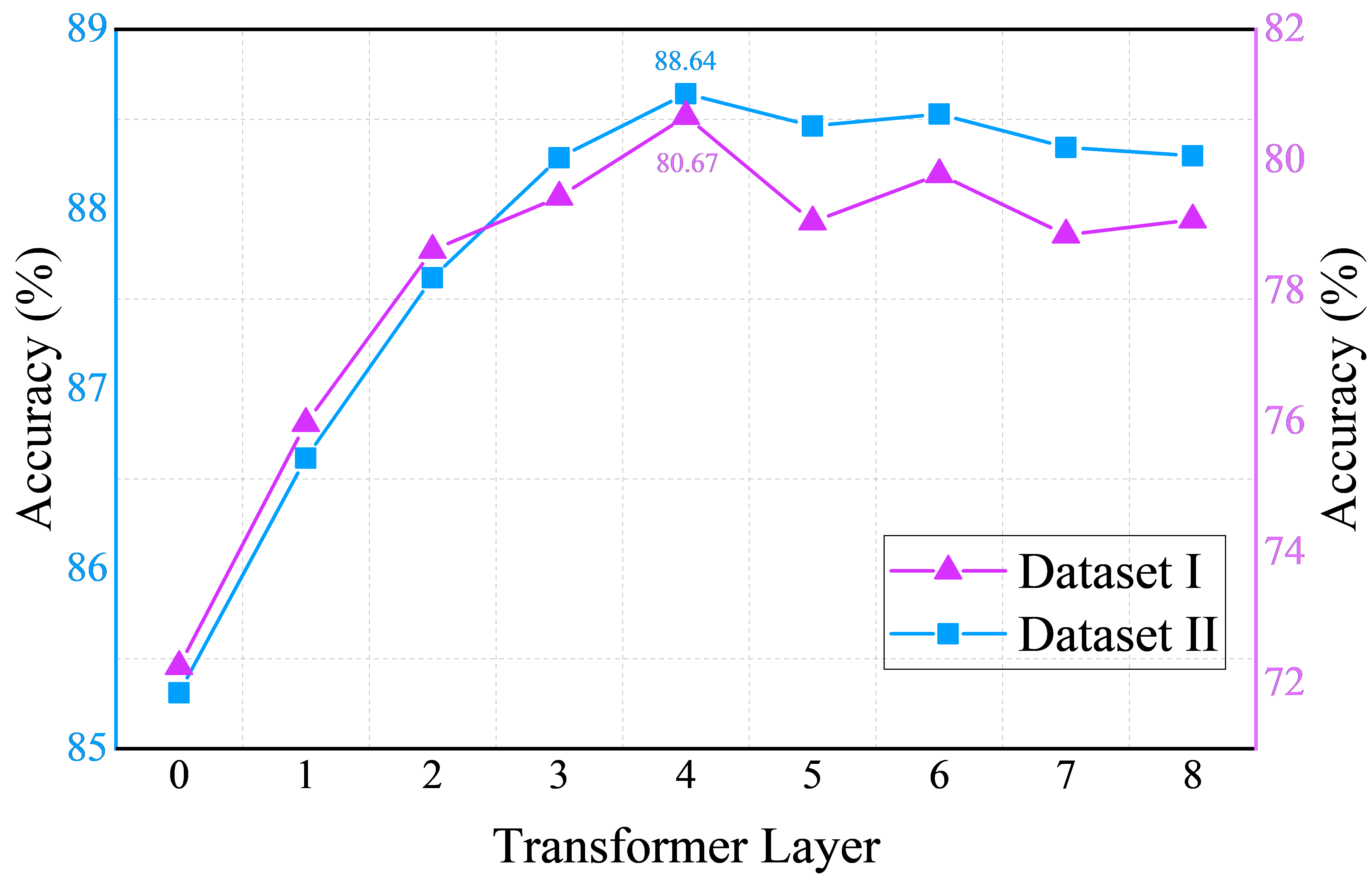}
\captionsetup{font=footnotesize}
\caption{The influence of the Transformer layer number on the average accuracy}
\vspace{-0.2cm}
\label{fig_5}
\end{figure}

Obviously, for DL models, the internal hyper-parameters of the network significantly affect its performance. The critical hyperparameters of our constructed model mainly include the dimensionality used for channel fusion and upscaling through pointwise convolution, the number of Transformer encoder layers and Transformer Heads.

First, here the pointwise dimension refers to the dimension parameter $D_2$, which is used in Dual-TSST for feature fusion and dimensionality increase through the pointwise convolution. To study the effects of this dimensional parameter, a range of [40, 160] with an interval of 10 has been designated, and Fig. \ref{fig_4} gives the resultant average accuracy. As shown in Fig. \ref{fig_4}, with an increase in $D_2$, the accuracy in Dataset I shows an overall trend of decreasing first, then rising, and finally maintaining a mild fluctuation. Similarly, the average accuracy corresponding to Dataset II initially increases with $D_2$ and then oscillates. Interestingly, the optimal dimensional parameter for both is found at $D_2$ = 120, which avoids the complexity associated with too high dimensions and effectively enhances the model expressive capabilities.

The number of Transformer layers refers to the stack levels of Transformer Encoders, which essentially define the depth of the model that determines the complexity and hierarchy of the information the model can learn. Generally, the deeper models typically enhance the model’s representational ability and fit the data better. However, as the number of layers increases, issues such as the overfitting and gradient explosion may occur, along with an increase in computational costs. The accuracy trends with changes in number of Transformer layers are illustrated in Fig \ref{fig_5}, where we see that the introduction of the Transformer (from zero to one layer) may lead to a marked performance improvement. Besides, while initial increases in layers (i.e., from 0 to 4) enhance performance for both datasets, the accuracy associated with all datasets begins to decline after the fourth layer. This may indicate that the model has reached its learning saturation or is beginning to overfit the noise within the data. For Dataset I, the peak accuracy with Dual-TSST exceeded the lowest by 4.36\% (p\textless0.01), and for Dataset II, the corresponding value is 2.03\% (p\textless0.05). These results suggest that while increasing the number of layers can enhance performance up to a certain limit, excessively high numbers may hinder training and increase the risk of overfitting.

\vspace{-0.5\baselineskip} 
\begin{figure}[h]
\centering
\captionsetup[subfloat]{font=scriptsize,labelformat = empty}
\subfloat[][]{\includegraphics[width = 8.1cm]{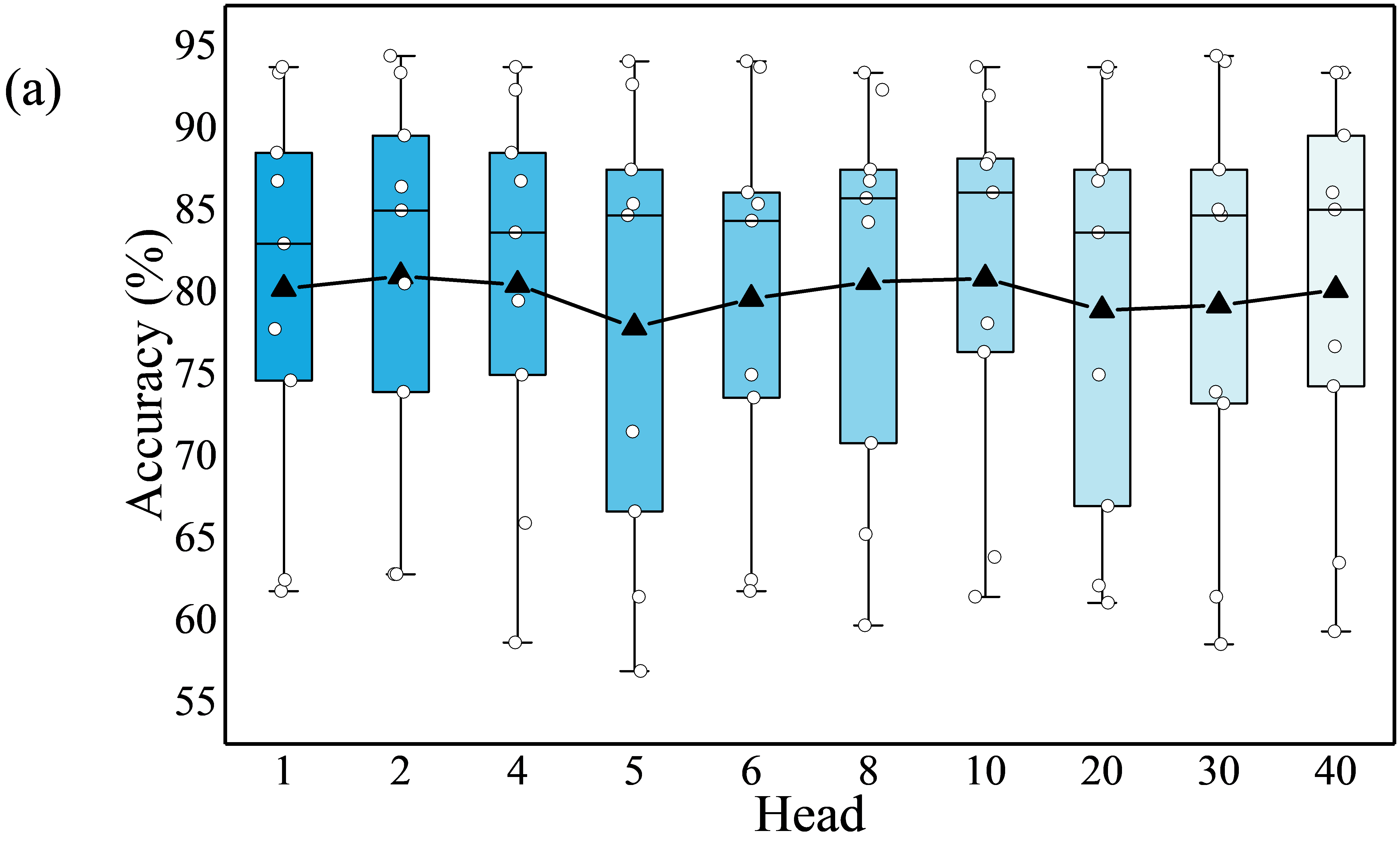}%
\label{fig6:subfig a}}\\
\vspace{-0.5cm}
\subfloat[][]{\includegraphics[width = 8.1cm]{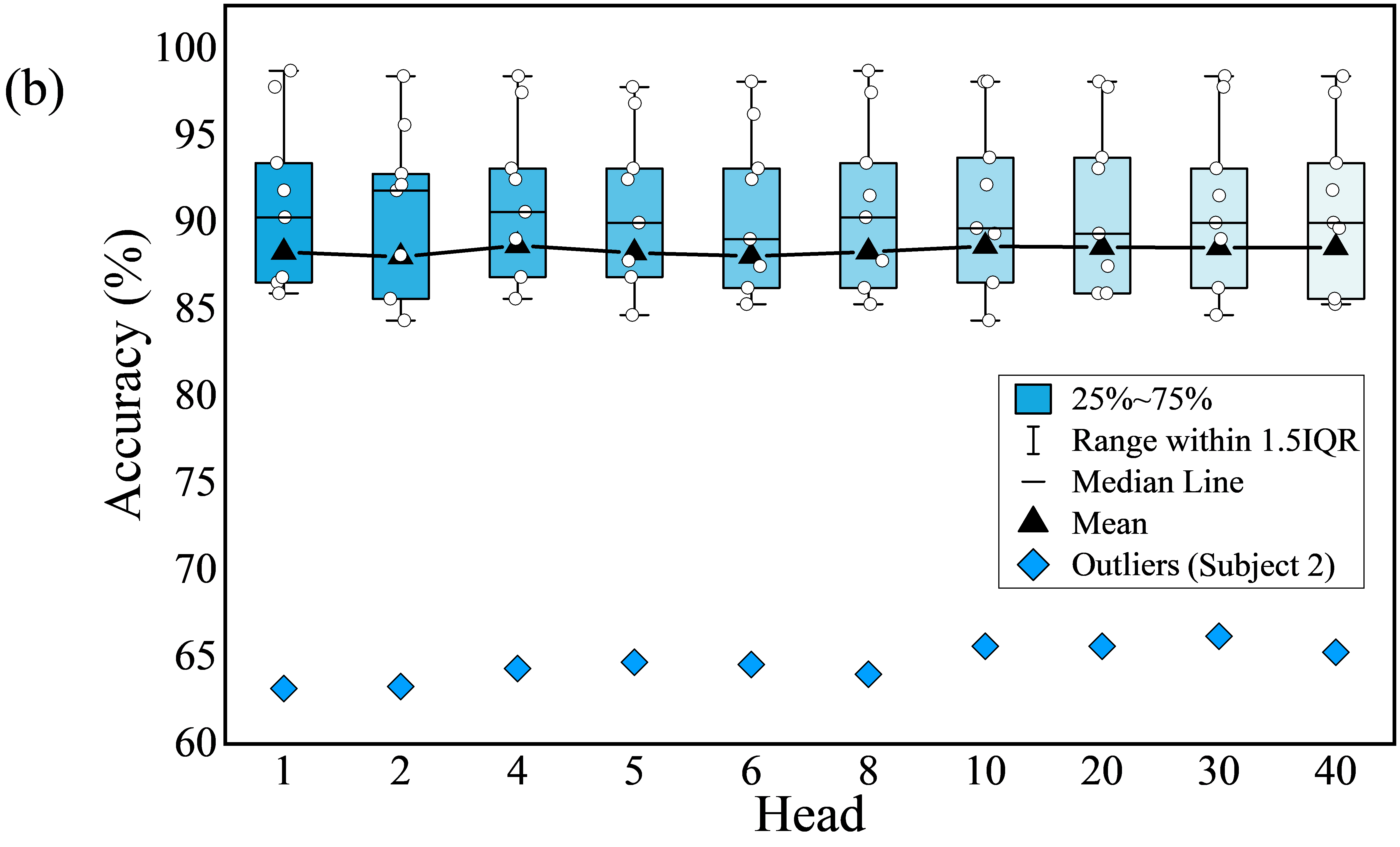}%
\label{fig6:subfig b}}
\captionsetup{font=footnotesize}
\vspace{-0.3cm}
\caption{The influence of the number of Multi-head attention heads on the accuracy for different datasets of (a) Dataset I, (b) Dataset II.}
\label{fig_6}
\end{figure}

In the Transformer model, each involved Head can be seen as an independent self-attention mechanism, while multi-head attention allows the model to concurrently attend to different semantic information, thereby capturing diverse relationships and features in the input sequence. More specifically, each head learns different weights to better encode information in various contexts, and thus enhancing the richness and expressive power of the representation. However, too many heads can also lead to overfitting or an increase in computational complexity. In this study, the influence of the heads number has been studied, for which the results are depicted in Fig. \ref{fig_6}. Noting that since the accuracy of Subject 2 of Dataset II is far from others, it is listed separately.

As illustrated in Fig. \ref{fig_6}, the average accuracy on Dataset I varies significantly with the number of heads, while on Dataset II, the fluctuation seems to be smaller. Overall, the highest accuracy for both Dataset I and II is achieved when the Head count was 10, showing an improvement to the lowest accuracy of 2.97\% (p\textgreater0.05) and 0.61\% (p\textgreater0.05), respectively.  Since the increase is not substantial, we conclude that changes in the number of Transformer heads do not significantly impact the model performance.

\vspace{-0.4cm}
\subsection{Ablation Experiment}

\vspace{-0.5\baselineskip} 
\begin{figure}[h]
\centering
\captionsetup[subfloat]{font=scriptsize,labelformat = empty}
\subfloat[][]{\includegraphics[width = 8.1cm]{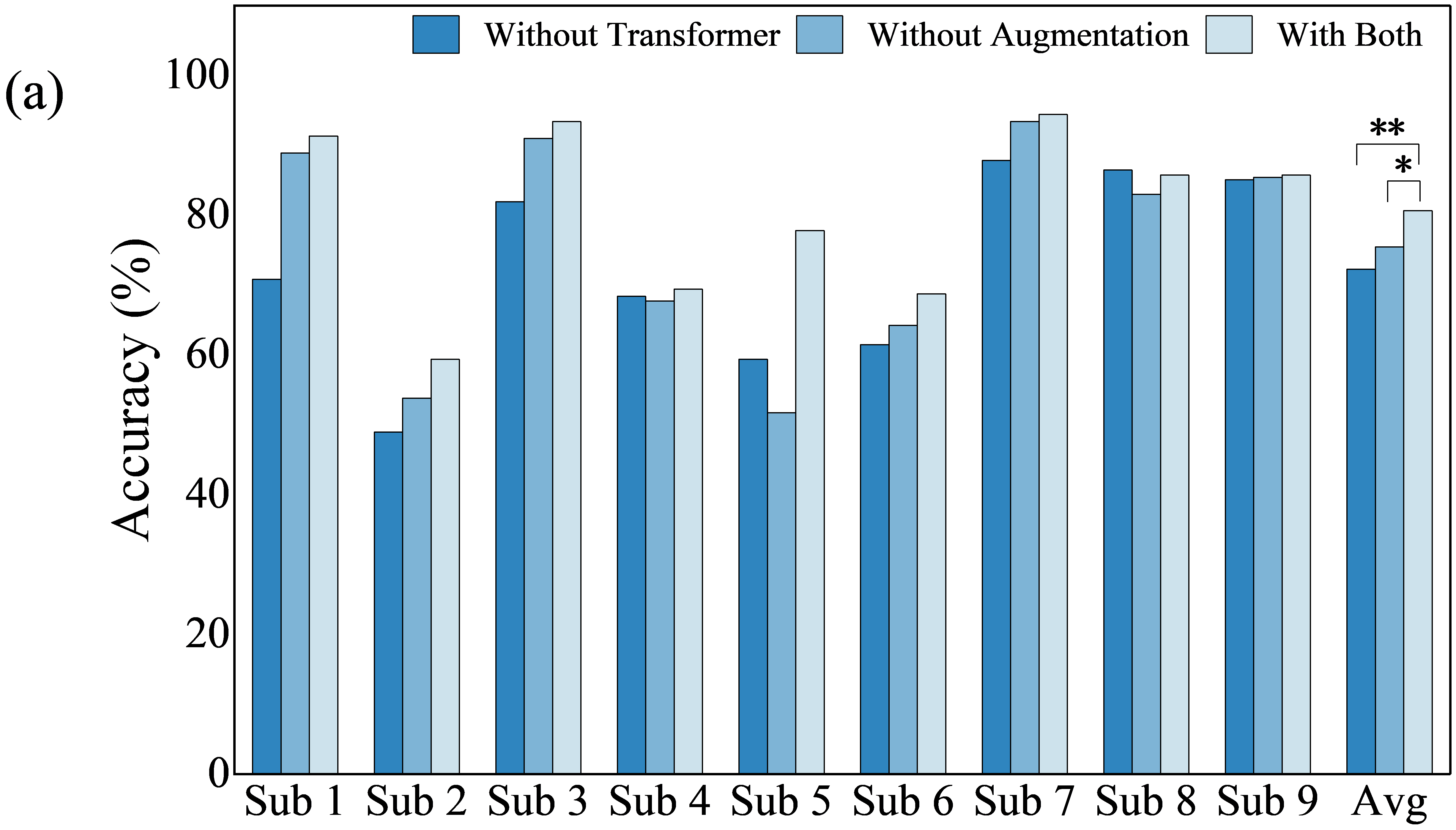}%
\label{fig7:subfig a}}\\
\vspace{-0.5cm}
\subfloat[][]{\includegraphics[width = 8.1cm]{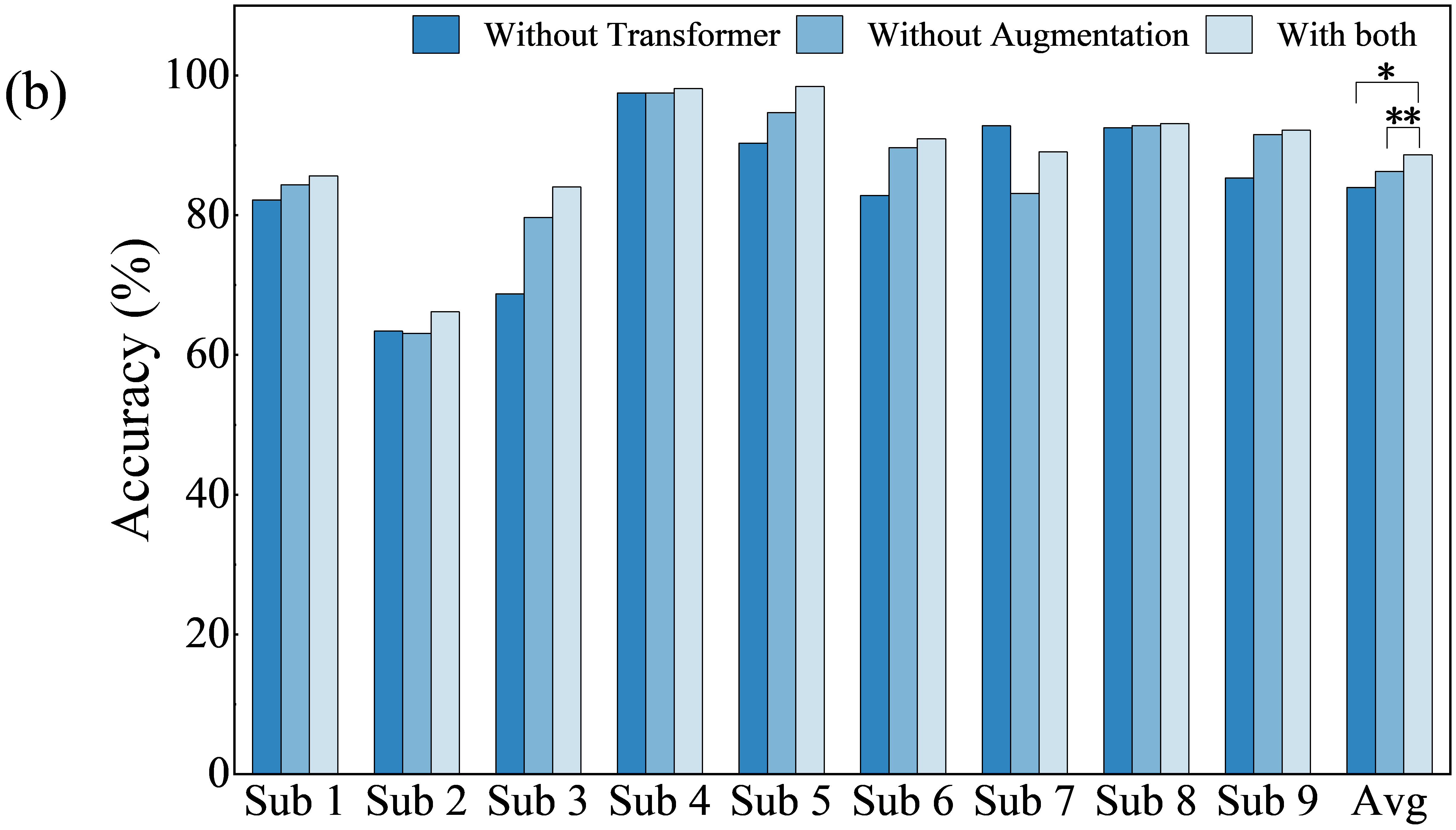}%
\label{fig7:subfig b}}
\captionsetup{font=footnotesize}
\vspace{-0.3cm}
\caption{Ablation experiments on data augmentation and Transformer application on datasets of (a) Dataset I, (b) Dataset II.}
\label{fig_7}
\end{figure}

The Dual-TSST model comprises multiple modules, and we introduced the data augmentation measures into the proposed framework. To determine the specific effects of each functional modules, ablation experiments were conducted on both Dataset I and Dataset II to assess the impact of data augmentation, the Transformer module, different branches, and various inputs.

Initially, we conducted ablation experiments on the data augmentation and Transformer modules. As illustrated in Fig. \ref{fig_7}, when it is without the Transformer module, we note an obvious decrease in the accuracy across most of the specific different subjects and the average results for the used datasets. However, an increase in performance is observed for Subject 7 of Dataset II, which possibly indicating overfitting when the module was used. Overall, for the tested two datasets, reintegrating the Transformer improved the overall average accuracy by 8.41\% (p\textless0.01) and 4.68\% (p\textless0.05), respectively, underscoring its critical role in boosting accuracy.

\begin{figure}[htpb]
\centering
\includegraphics[width=7.62cm]{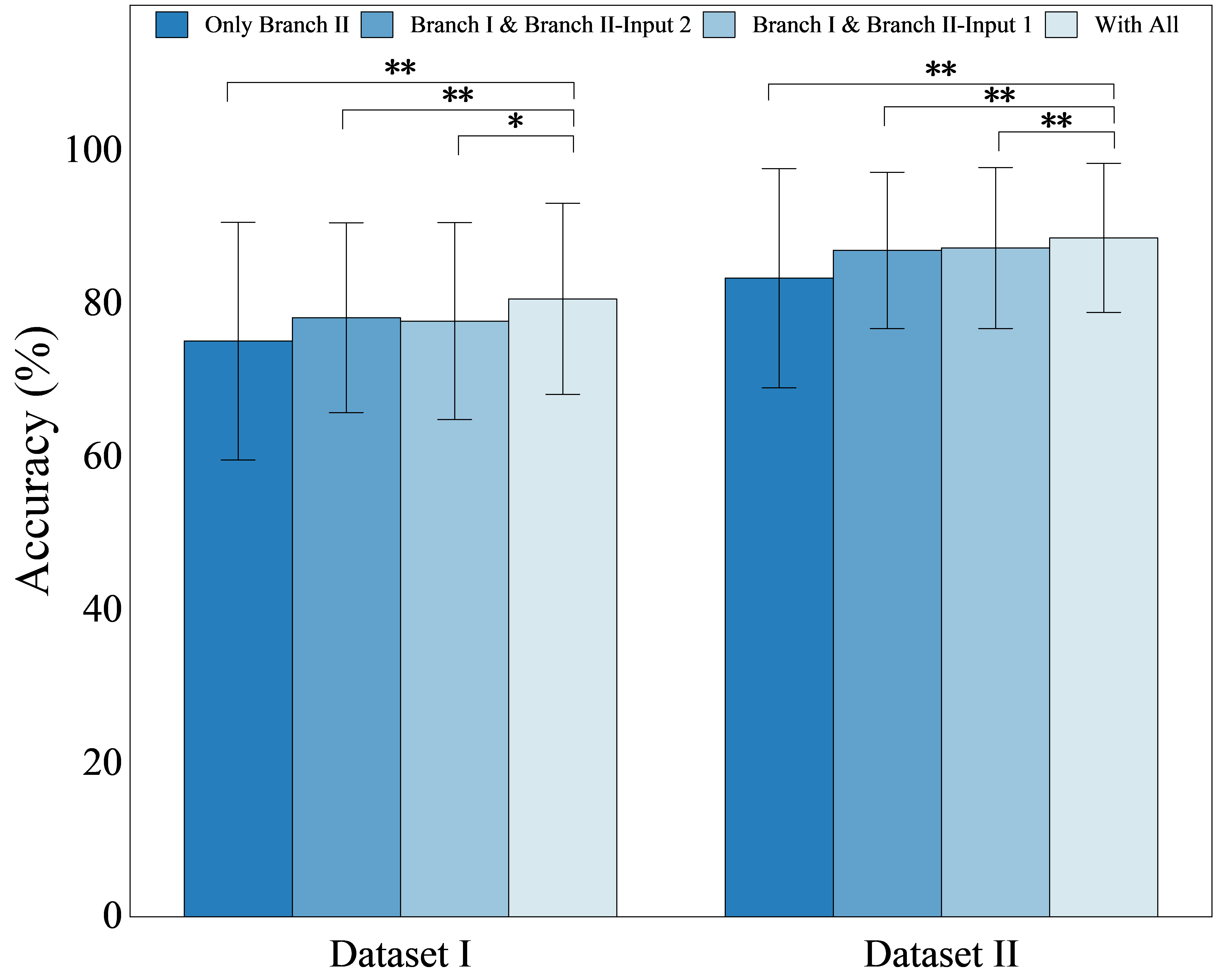}
\captionsetup{font=footnotesize}
\caption{Ablation experiments on different branches and inputs on both datasets.}
\vspace{-0.5cm} 
\label{fig_8}
\end{figure}

\begin{figure*}[tb]
\centering
\includegraphics[width=18cm]{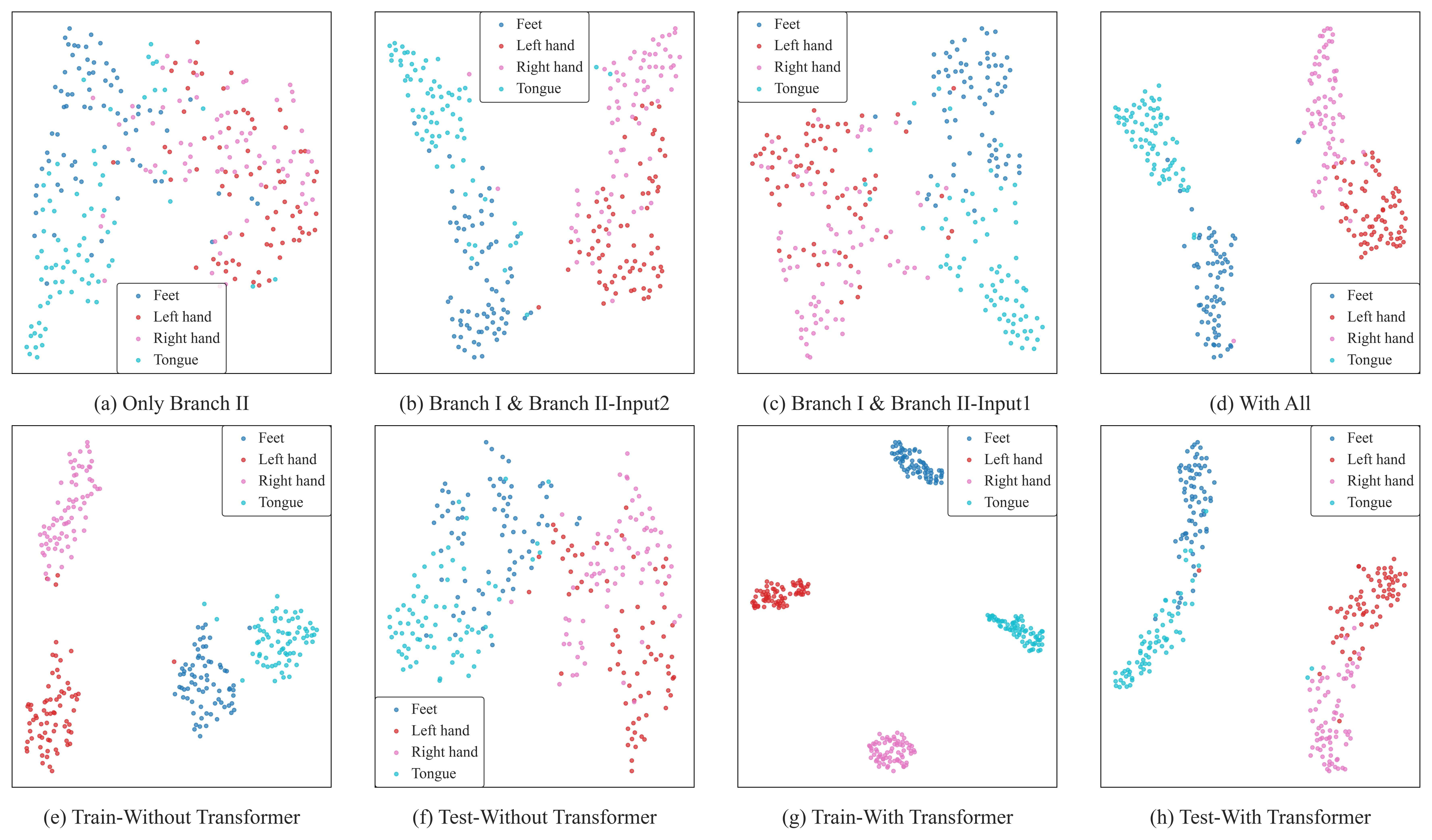}
\captionsetup{font=footnotesize}
\caption{The comparison of the features for Subject 7 of Dataset I of with/without the Transformer module by t-SNE visualization.}
\vspace{-0.2cm}
\label{fig_9}
\end{figure*}

The operation of data augmentation is envisioned to expand the data scale, aid the model in capturing more complex patterns, and mitigate overfitting tendencies. Meanwhile, it also introduces additional variability and disturbances. Across the datasets I and II, the application of data augmentation strategies led to a 5.21\% (p\textless0.05) and 2.37\% (p\textless0.01) increase in average accuracy, which indicates that such the module has proven to enhance model performance significantly.

We further conducted experiments by removing Branch I and Branch II (with Input 1 or Input 2), where the results of the remaining parts are reported in Fig. \ref{fig_8}. It was observed that removing Branch I (i.e., only Branch II) significantly impacted the overall performance on both datasets (p\textless0.05), because Branch I provides the majority of the temporal features to the model. Besides, removing the input from Branch II had some impact, but not as significant as that from Branch I. Overall, on both datasets, performance using two branches was superior to using just one. Within Branch II, using two inputs also showed an improvement over using just one input. In addition, the improved error bar range of the model with all branches implies the enhanced robustness.

\vspace{-0.4cm}
\subsection{Visualization}

To further intuitively demonstrate the effectiveness of the designed branches and self-attention mechanism, a compara-tive study of low-dimensional visualizations, using t-SNE \cite{48}, was conducted for one typical subject (i.e., Subject 7 of Dataset I). Fig. \ref{fig_9} reports the relevant results with/without prominent components (e.g., Branch I or Branch II of feature extraction part, Transformer modules). Specifically, for the test data, as in Fig. 9 (a), when it with Branch II only, the features of focused categories are closely mixed. In contrast, as shown in (b) and (c), the distance between classes becomes larger with the help of Branch I, even if only a part of Branch II is involved. The results of inter-category distance is being more evident with all developed branches, thus illustrating the capacity of our model.

Moreover, as we can see from Fig. \ref{fig_9} (e), without the Transformer, the t-SNE visualization of the training set reveals several well-separated clusters, indicating a clear distribution of categories in low-dimensional space. However, when applied to the testing set, the model exhibits a significant reduction in category separation (see Fig. \ref{fig_9} (f)). In particular, a substantial overlap of features between the feet and tongue, left- and right-hands can be apparently observed. This overlap indicates that while efficiently learning the properties of each category on the training data, the model exhibits poor generalization when exposed to unknown data, failing to discriminate between comparable classes. Conversely, the introduction of Transfor¬mer results in a dramatic improvement. As in t-SNE visualiza¬tion, the training set displays highly distinct and well-separated clusters, with each category occupying a clear, even non-overlapping region in low-dimensional space. This implies that the transformer module significantly improves the model’s capacity to describe diverse properties, resulting in a better defined distribution of categories. Importantly, the t-SNE visualization of the test set also exhibits considerable improvement, where the distinctions between hand features and other categories become more prominent. Especially in categories prone to confusion (such as left and right hand), the Transformer module significantly lowers overlap, highlighting its vital role in strengthening the model’s generalization performance and capacity to differentiate between comparable categories.

\begin{figure}[htpb]
\centering
\includegraphics[width=8.89cm]{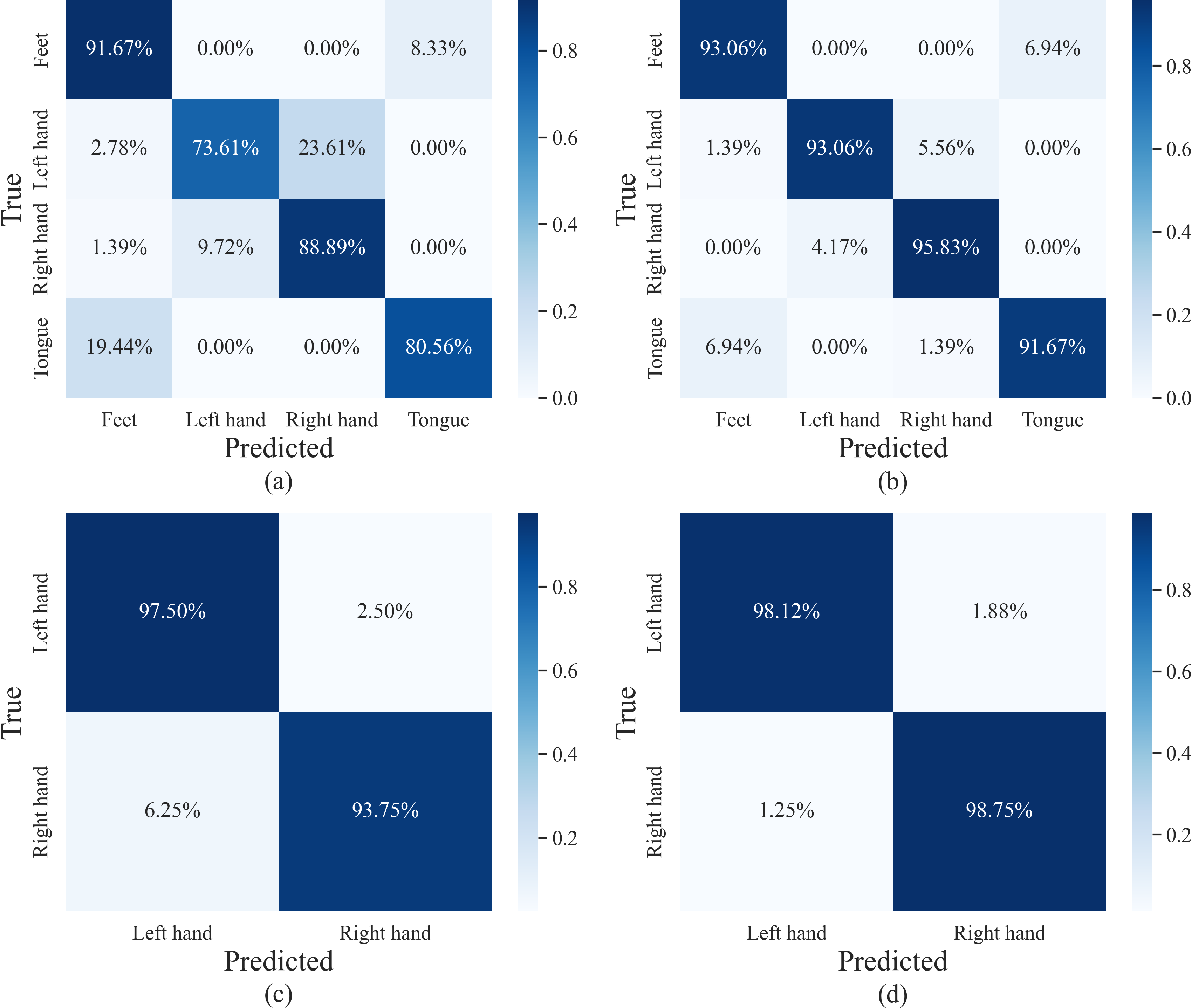}
\captionsetup{font=footnotesize}
\caption{Confusion matrices for (a) Subject 7 of Dataset I without Transformer, (b) Subject 7 of Dataset I with Transformer, (c) Subject 4 of Dataset II without Transformer, (d) Subject 4 of Dataset II with Transformer. }
\vspace{-0.5cm}
\label{fig_10}
\end{figure}

To further exhibit the impact of the integrated Transformer modules, graphical confusion matrix was used to present the classification performance across the specific categories. For each dataset, the results of one subject with/without related part are depicted in Fig. \ref{fig_10}. As it can be seen, the results clearly demonstrate that the model without the Transformer module

faces considerable challenges in capturing the discriminative features. For instance, the confusion matrix reveals that 23.61\% of left-hand features were erroneously classified as right-hand features, whereas a notable 19.44\% misclassification rate of tongue features being recognized as feet (see in Fig. \ref{fig_10} (a)). Such results suggest that the model struggles with discerning subtle feature differences, which may lead to a generalization inadequacy, particularly when dealing with categories that exhibit similar features. Instead, upon incorporating the Trans¬former, a marked improvement in the model’s classification capabilities is observed (mere the value of 5.56\% and 6.94\% of corresponding index are found). Also for hand imaginary recognition of Fig. \ref{fig_10} (c) and (d), following the implementation of the Transformer, the model for subject 4 of dataset II has a particularly satisfactory classification accuracy of 98.75\%, a notable increase of near 5\%. These improvements suggests that the Transformer module bolsters the model’s feature extraction capabilities and enhances its generalization ability and robust¬ness across different datasets.

\section{Disscussion and Conclusion}
\label{chapt:5}
The statistical distribution of non-stationary EEG data varies across different subjects and recording sessions, making it challenging for BCI researchers to design a classifier with high accuracy and generalization capability. Borrowing the idea of machine learning with data flow of feature extraction, feature fusion, and classification, a novel efficient DL-framework that fully incorporates the CNN and Transformer is suggested in this study to handle the data processing of EEG signals.

The proposed dual-TSST model first leverages a dual-branch CNN structure, which accepts data from diverse perspectives, to dig the comprehensive representation of entailed features. In this configuration, Branch I is tasked with extracting spatio¬temporal features from raw EEG, while Branch II handles the spatio-temporal-frequency features from wavelet-transformed data. The Transformer module further explores the long-range global dependencies and synthesizes all the diverse features into a cohesive feature set, which is finally classified by the classification module. For the proposed dual-TSST framework, minimal yet critical preprocessing with band-pass filtering, as in \cite{16,26}, is needed to EEG signals, which avoids the specific sophisticated data preprocessing steps. In essence, such the proposed DL model does not require the extra expert knowledge but can automatically extract the comprehensive spatio-temporal-frequency features, which is conducive to the identification from multifaceted data sources.

Experimentally, the framework was evaluated through two BCI Competition IV Datasets of 2a, 2b, and one widely used emotional Dataset of SEED, where the superior performance compared with state-of-the-art methods has been achieved. In general, extensive parameter sensitivity and ablation studies affirm that each component significantly contributes to the model’s effectiveness, particularly highlighting the substantial impact of pointwise dimension and Transformer. Particularly, the number of Transformer layers, which also being termed as the depth in other related studies, directly influences the model’s classification result, highlighting the importance of such the module introduction. More importantly, our specific results of Fig. 5 reveal the instructive suggestion for subsequent layer configuration of future transformer-based EEG decoding. Conversely, the number of heads in the multi-head attention setup showed marginally impact on the final performance, and such insensitivity may aid in the lightweight iterative design of future models. Moreover, the effects of the related branches and data augmentation have also been intuitively presented, thus clarifying the rationality of the developed framework. 

Whereas the approach proposed in this study boosts the model capability to extract more discriminative EEG features, it still has several limitations. First, a notable limitation of our current model is its structural complexity. The majority of the parameters in current model originates from the comprehensive Transformer module and the fully connected layers for the classification, which coincides with the prior research of EEG Conformer \cite{16}. Although depthwise separable convolutions were applied to mitigate this issue, current model still maintains a higher parameter count. Second, as the deep learning model, the number of samples for the focused data is expected to be large enough. While the data augmentation strategies can be strategically adopted as current work, one should maintain the quest for more effective source data \cite{40}. Since it is expensive, not-friendly, and impractical to always collect a larger number of recording data, several advanced methods, such as the transfer learning based domain adaptation, which uses knowledge from source subjects to improve the performance of a targeted one \cite{38}, ought to be applied toward all accessible data. Moreover, only the subject-specific based experiments are conducted in this study, while more cross-subjects validation should be focused to further investigate the generalizability of the model \cite{17,26}. 

To sum up, the developed innovative architecture leverages the distinct strengths of related data types to enhance the accuracy and robustness of the decoding process, while also improving network interpretability through obeying ML-based processing flow. Moving forward, our objectives will focus on optimizing the model’s architecture and reducing its parameter footprint, alongside exploring online potential applications.


\begin{thebibliography}{50}

\bibitem{1}
J. R. Wolpaw \textit{et al.}, “Brain-computer interface technology: a review of the first international meeting,”  \textit{IEEE Trans. Neural Syst. Rehabil. Eng.}, vol. 8, no. 2, pp. 164-173, Jun. 2000.
\bibitem{2}
H. Li, L. Bi, X. Li, and H. Gan, “Robust predictive control for EEG-based brain–robot teleoperation,”  \textit{IEEE Trans. Intell. Transp. Syst.}, vol. 25, no. 8, pp. 9130-9140, Aug. 2024.
\bibitem{3}
H. Li, L. Bi, and J. Yi, “Sliding-mode nonlinear predictive control of brain-controlled mobile robots,”  \textit{IEEE Trans. Cybern.}, vol. 52, no. 6, pp. 5419-5431, Jun. 2022.
\bibitem{4}
H. Li, L. Bi, and H. Shi, “Modeling of human operator behavior for brain-actuated mobile robots steering,”  \textit{IEEE Trans. Neural. Syst. Rehabil. Eng.}, vol. 28, no. 9, pp. 2063-2072, Sep. 2020.
\bibitem{5}
R. Abiri, S. Borhani, E. W. Sellers, Y. Jiang, and X. Zhao, “A comprehensive review of EEG-based brain–computer interface paradigms,”  \textit{J. Neural Eng.}, vol. 16, no. 1, Feb. 2019, Art. no. 011001.
\bibitem{6}
S. Aggarwal and N. Chugh, “Review of machine learning techniques for EEG based brain computer interface,”  \textit{Arch. Comput. Method Eng.}, vol. 29, no. 5, pp. 3001-3020, Aug. 2022.
\bibitem{7}
S. Gong, K. Xing, A. Cichocki, and J. Li, “Deep learning in EEG: advance of the last ten-year critical period,”  \textit{IEEE Trans. Cognit. Develop. Syst.}, vol. 14, no. 2, pp. 348-365, Jun. 2022.
\bibitem{8}
Y. LeCun, Y. Bengio, G. Hinton, “Deep learning,”  \textit{Nature}, vol. 521, no. 7553, pp. 436-444, May. 2015.
\bibitem{9}
Z. Li, F. Liu, W. Yang, S. Peng, and J. Zhou, “A survey of convolutional neural networks: analysis, applications, and prospects,”  \textit{IEEE Trans Neural Netw. Learn. Syst.}, vol. 33, np. 12, pp. 6999-7019, Dec. 2022.
\bibitem{10}
W. Rawat and Z. Wang, “Deep convolutional neural networks for image classification: A comprehensive review,”  \textit{Neural Comput.}, vol. 29, no. 9, pp. 2352-2449, Sep. 2017.
\bibitem{11}
D. W. Otter, J. R. Medina, and J. K. Kalita, “A survey of the usages of deep learning for natural language processing,”  \textit{IEEE Trans Neural Netw. Learn. Syst.}, vol. 32, no. 2, pp. 604-624, Feb. 2021.
\bibitem{12}
R. T. Schirrmeister  \textit{et al.}, “Deep learning with convolutional neural networks for EEG decoding and visualization,”  \textit{Human. Brain Mapp.}, vol. 38, no. 11, pp. 5391-5420, Aug. 2017.
\bibitem{13}
V. J Lawhern  \textit{et al.}, “EEGNet: a compact convolutional neural network for EEG-based brain–computer interfaces,”  \textit{J. Neural Eng.}, vol. 15, no. 5, Jul. 2018, Art. no. 056013.
\bibitem{14}
[S. Tortora  \textit{et al.}, “Deep learning-based BCI for gait decoding from EEG with LSTM recurrent neural network,”  \textit{J. Neural Eng.}, vol. 17, no. 4, Jul. 2020, Art. no. 046011.
\bibitem{15}
J. Sun, J. Xie, and H. Zhou, “EEG classification with transformer-based models,” in \textit{ Proc. IEEE 3rd Glob. Conf. Life Sci. Technol., (LifeTech).}, pp. 92-93, 2021.
\bibitem{16}
Y. Song, Q. Zheng, B. Liu, and X. Gao, “EEG conformer: convolutional transformer for EEG decoding and visualization,”  \textit{IEEE Trans. Neural Syst. Rehabil. Eng.}, vol. 31, pp. 710-719, Dec. 2023.
\bibitem{17}
W. Tao  \textit{et al.}, “ADFCNN: attention-based dual-scale fusion convolu- tional neural network for motor imagery brain–computer interface,”  \textit{IEEE Trans. Neural Syst. Rehabil. Eng.}, vol. 32, pp. 154-165. 2024.
\bibitem{18}
A. Arjun, A. S. Rajpoot, and M. Raveendranatha Panicker, “Introducing attention mechanism for EEG signals: emotion recognition with vision transformers,” in  \textit{Proc. 43rd Annu. Int. Conf. IEEE Eng. Med. Biol. Soc. (EMBC).}, pp. 5723-5726, Nov. 2021.
\bibitem{19}
M. S. Al-Quraishi  \textit{et al}., “Decoding the user’s movements preparation from EEG signals using vision transformer architecture,”  \textit{IEEE Access}, vol. 10, pp. 109446-109459, Oct. 2022.
\bibitem{20}
M. A. Mulkey  \textit{et al.}, “Supervised deep learning with vision transformer predicts delirium using limited lead EEG,”  \textit{Sci. Rep.}, vol. 13, no. 1, May. 2023, Art. no. 7890.
\bibitem{21}
A. Nogales  \textit{et al}., “BERT learns from electroencephalograms about Parkinson’s disease: transformer-based models for aid diagnosis,”  \textit{IEEE Access}, vol. 10, pp. 101672-101682, Jan. 2022.
\bibitem{22}
B. Wang, X. Fu, Y. Lan, L. Zhang, and Y. Xiang, “Large transformers are better EEG learners,” \textit{arXiv: 2308.11654}.
\bibitem{23}
J. Zhou, Y. Duan, Y. Zou, Y. -C. Chang, Y. -K. Wang, and C. -T. Lin, “Speech2EEG: leveraging pretrained speech model for EEG signal recognition,”  \textit{IEEE Trans. Neural Syst. Rehabil. Eng.}, vol. 31, pp. 2140-2153, Apr. 2023.
\bibitem{24}
X. Tian  \textit{et al.}, “Deep multi-view feature learning for EEG-based epileptic seizure detection,”  \textit{IEEE Trans. Neural Syst. Rehabil. Eng.}, vol. 27, no. 10, pp. 1962-1972, Oct. 2019.
\bibitem{25}
R. Mane  \textit{et al.}, “A multi-view CNN with novel variance layer for motor imagery brain computer interface,” in  \textit{Proc. 42nd Annu. Int. Conf. IEEE Eng. Med. Biol. Soc. (EMBC).}, pp. 2950-2953, Jul. 2020.
\bibitem{26}
H. Zhi, Z. Yu, T. Yu, Z. Gu, and J. Yang, “A multi-domain convolutional neural network for EEG-based motor imagery decoding,”  \textit{IEEE Trans. Neural Syst. Rehabil. Eng.}, vol. 31, pp. 3988-3998, Oct. 2023.
\bibitem{27}
G. Liang, D. Cao, J. Wang, Z. Zhang, and Y. Wu, “EISATC-fusion: inception self-attention temporal convolutional network fusion for motor imagery EEG decoding,”  \textit{IEEE Trans. Neural Syst. Rehabil. Eng.}, vol. 32, pp. 1535-1545, Mar. 2024.
\bibitem{28}
Y. Qin, B. Yang, S. Ke, P. Liu, F. Rong, and X. Xia, “M-FANet: multi-feature attention convolutional neural network for motor imagery decoding,”  \textit{IEEE Trans. Neural Syst. Rehabil. Eng.}, vol. 32, pp. 401-411, Jan. 2024.
\bibitem{29}
C. Liu  \textit{et al.}, “SincNet-based hybrid neural network for motor imagery EEG decoding,”  \textit{IEEE Trans. Neural Syst. Rehabil. Eng.}, vol. 30, pp. 540-549, Mar. 2022.
\bibitem{30}
M. X. Cohen, “A better way to define and describe Morlet wavelets for time-frequency analysis,”  \textit{NeuroImage}, vol. 199, pp. 81-86, Oct. 2019.
\bibitem{31}
F. Lotte, “Signal processing approaches to minimize or suppress calibra¬tion time in oscillatory activity-based brain–computer interfaces,”  \textit{Proc IEEE}, vol. 103, no. 6, pp. 871-890, Jun. 2015.
\bibitem{32}
M. Tangermann  \textit{et al.}, “Review of the BCI competition IV,”  \textit{Front. Neurosci.}, vol. 6, p.55, Jul. 2012.
\bibitem{33}
V. Jayaram and A. Barachant, “MOABB: Trustworthy algorithm bench¬marking for BCIs,”  \textit{J. Neural Eng.}, vol. 15, no. 6, Dec. 2018, Art. no. 066011.
\bibitem{34}
W. L. Zheng and B. L. Lu, “Investigating critical frequency bands and channels for EEG-based emotion recognition with deep neural networks,”  \textit{IEEE Trans. Auton. Mental Develop.}, vol. 7, no. 3, pp. 162-175, Sep. 2015.
\bibitem{35}
I. Loshchilov and F. Hutter, “SGDR: Stochastic gradient descent with warm restarts,” 2016,  \textit{arXiv:1608.03983}.
\bibitem{36}
R. Mane,  \textit{et al}. “FBCNet: A multi-view convolutional neural network for brain-computer interface,” 2021,  \textit{arXiv:2104.01233}.
\bibitem{37}
H. Zhao, Q. Zheng, K. Ma, H. Li, and Y. Zheng, “Deep representation¬based domain adaptation for nonstationary EEG classification,”  \textit{IEEE Trans Neural Netw. Learn. Syst.}, vol. 32, no. 2, pp. 535-545, Feb. 2021.
\bibitem{38}
C. Phunruangsakao, D. Achanccaray, and M. Hayashibe, “Deep adversa¬rial domain adaptation with few-shot learning for motor-imagery brain¬computer interface,”  \textit{IEEE Access}, vol. 10, pp. 57255-57265, Jan. 2022.
\bibitem{39}
A. Salami, J. Andreu-Perez, and H. Gillmeister, “EEG-ITNet: an explainable inception temporal convolutional network for motor imagery classification,”  \textit{IEEE Access}, vol. 10, pp. 36672-36685, Apr. 2022.
\bibitem{40}
J. Wang, L. Yao, and Y. Wang, “IFNet: An interactive frequency convolutional neural network for enhancing motor imagery decoding from EEG,”  \textit{IEEE Trans. Neural Syst. Rehabil. Eng.}, vol. 31, pp. 1900-1911, Jan. 2023.
\bibitem{41}
P. Chen, Z. Gao, M. Yin, J. Wu, K. Ma, and C. Grebogi, “Multiattention adaptation network for motor imagery recognition,”  \textit{IEEE Trans. Syst. Man, Cybern. Syst.}, vol. 52, no. 8, pp. 5127-5139, Aug. 2022.
\bibitem{42}
X. Tang, C. Yang, X. Sun, M. Zou, and H. Wang, “Motor imagery EEG decoding based on multi-scale hybrid networks and feature enhance¬ment,”  \textit{IEEE Trans. Neural Syst. Rehabil. Eng.}, vol. 31, pp. 1208-1218, Feb. 2023.
\bibitem{43}
M. Jiménez-Guarneros and G. Fuentes-Pineda, “Cross-subject EEG- based emotion recognition via semisupervised multisource joint distribu¬tion adaptation,”  \textit{IEEE Trans. Instrum. Meas.}, vol. 72, pp. 1–12, 2023.
\bibitem{44}
Y. Li  \textit{et al.}, “A novel Bi-hemispheric discrepancy model for EEG emotion recognition,”  \textit{IEEE Trans. Cogn. Develop. Syst.}, vol. 13, no. 2, pp. 354-367, 2021.
\bibitem{45}
P. Zhong, D. Wang and C. Miao, “EEG-based emotion recognition using regularized graph neural networks,”  \textit{IEEE Trans. Affect. Comput.}, vol. 13, no. 3, pp. 1290-1301, 2022.
\bibitem{46}
L. Yang  \textit{et al.}, “Electroencephalogram-based emotion recognition using factorization temporal separable convolution network,”  \textit{Eng. Appl. Artif. Intell.}, vol. 133, 2024, Art. no. 108011.
\bibitem{47}
J. Liu  \textit{et al.}, “Spatial-temporal transformers for EEG emotion recognition,” in  \textit{Proc. Int. Conf. Adv. Artif. Intell.}, 2022, pp. 116-120. 
\bibitem{48}
L. Van der Maaten and G. Hinton, “Visualizing data using t-SNE,”  \textit{J. Mach. Learn. Res.}, vol. 9, no. 11, pp. 2579-2605, 2008.


\end{thebibliography}
\end{document}